\documentclass[]{aa}
\usepackage[varg]{txfonts}
\usepackage{natbib}
\usepackage{latexsym,amsmath,amsfonts,amssymb,textcomp}
\usepackage{graphicx}

\begin{document}

\title{The habitability of a stagnant-lid Earth}

\author{N. Tosi\inst{1,2} \and M. Godolt\inst{1,2} \and B. Stracke\inst{2} \and T. Ruedas\inst{3,2} \and J. L.
Grenfell\inst{2} \and D. H\"oning\inst{2} \and \\ A. Nikolaou\inst{1,2}  \and A.-C. Plesa\inst{2} \and D. Breuer\inst{2} \and T. Spohn\inst{2}}

\institute{Technische Universit\"at Berlin, Department of Astronomy and Astrophysics, Hardenbergstr. 36, 10623
Berlin, Germany \and Institute for Planetary Research, German Aerospace Center (DLR), Rutherfordstr. 2, 12489
Berlin, Germany,  \and Westf\"alische Wilhelms-Universit\"at M\"unster, Institute of Planetology, Wilhelm-Klemm-Str. 10,
48149 M\"unster, Germany}

\date{}

\abstract 
%
%
{Plate tectonics is considered a fundamental component for the habitability of the Earth. Yet whether it is a recurrent feature of terrestrial bodies orbiting other stars or unique to the Earth is unknown. The stagnant lid may rather be the most common tectonic expression on such bodies.} 
%
%
{To understand whether a stagnant-lid planet can be habitable, i.e. host liquid water at its surface, we model the thermal evolution of the mantle, volcanic outgassing of H$_2$O and CO$_2$, and resulting climate of an Earth-like planet lacking plate tectonics.}
%
%
{We used a 1D model of parameterized convection to simulate the evolution of melt generation and the build-up of an atmosphere of H$_2$O and CO$_2$ over 4.5 Gyr. We then employed a 1D radiative-convective atmosphere model to calculate the global mean atmospheric temperature and the boundaries of the habitable zone (HZ).} 
%
%
{The evolution of the interior is characterized by the initial production of a large amount of partial melt accompanied by a rapid outgassing of H$_2$O and CO$_2$. The maximal partial pressure of H$_2$O is limited to a few tens of bars by the high solubility of water in basaltic melts. The low solubility of CO$_2$ instead causes most of the carbon to be outgassed, with partial pressures that vary from 1 bar or less if reducing conditions are assumed for the mantle to 100--200 bar for oxidizing conditions. At 1 au, the obtained temperatures generally allow for liquid water on the surface nearly over the entire evolution. While the outer edge of the HZ is mostly influenced by the amount of outgassed CO$_2$, the inner edge presents a more complex behaviour that is dependent on the partial pressures of both gases.} 
%
%
{At 1 au, the stagnant-lid planet considered would be regarded as habitable. The width of the HZ at the end of the evolution, albeit influenced by the amount of outgassed CO$_2$, can vary in a non-monotonic way depending on the extent of the outgassed H$_2$O reservoir. Our results suggest  that stagnant-lid planets can be habitable  over geological timescales and that joint modelling of interior evolution, volcanic outgassing, and accompanying climate is necessary to robustly characterize planetary habitability.}
\maketitle

\section{Introduction}\label{sec_intro}

One of the central issues regarding the potential habitability of extrasolar planets is the extent to which plate tectonics is required to maintain habitability \citep{southam2015}. Plate tectonics is considered to be crucial for maintaining the activity of the carbon-silicate cycle over geological timescales. This helps to stabilize the climate and hence contributes to the habitability of Earth and possibly other planets \citep[e.g.][]{walker1981,kasting1993a}. Indeed, the boundaries of the habitable zone (HZ), i.e. the region surrounding a star where liquid water can be stable on a planetary surface, are traditionally calculated under the assumption that plate tectonics operates effectively \citep{kasting1993a,kopparapu2014}.  

On the one hand, even the Earth, where plate tectonics and a carbon--silicate cycle are active, could become
uninhabitable by turning into a snowball if the degassing rate of CO$_2$ from the interior becomes too low
\citep{tajika2007,kadoya2014,kadoya2015}. On the other hand, the very way in which plate tectonics operates on
Earth, when it started, and whether it is a stable or a transient feature in the tectonic history of our planet are
all complex and still controversial matters
\citep[e.g.][]{tackley2000a,bercovici2003a,bercovici2003b,vanhunen2008,vanhunen2012,gerya2014,oneill2016}. For the
numerous super-Earths --   large terrestrial extrasolar planets with masses between 1 and 10 $M_\oplus$ -- which have been detected in the past few years \citep[e.g.][]{batalha2014}, it has proven difficult to establish only on the basis of mass and radius whether plate tectonics is more or less likely to occur than on Earth. While some authors argue for a reduced tendency for plate tectonics to take place on these bodies  \citep{oneill2007c,kite2009,stamenkovic2012,stein2013}, others favour an increased tendency \citep{valencia2007, vanheck2011, orourke2012} or suggest that the tectonic behaviour of a rocky body can be strongly affected by the specific thermal conditions present after planetary formation and by the particular thermochemical history experienced by the interior \citep[e.g.][]{noack2014b,oneill2016}.

Even if an Earth twin with the same mass and radius as our planet and at the same distance from a Sun-like star was detected, it would be very difficult to establish whether plate tectonics exists on such a body or not; finding rocky planets
in the HZ is one of the major goals of exoplanetary missions such as the
upcoming PLATO 2.0 \citep{rauer2014}. In the absence of more detailed information and based on the limited evidence of the solar system, the stagnant lid may conservatively be considered as the most common tectonic mode under which terrestrial bodies operate. Because of the strong exponential dependence of mantle viscosity on temperature, the relatively cold upper layers of a rocky body naturally tend to be highly stiff and form a single, immobile plate, i.e. a stagnant lid. Contrary to tectonic plates, this lid does not participate in the dynamics of the mantle and does not allow surface materials to be directly recycled into the deep interior \citep[e.g.,][]{christensen1984,davaille1993,solomatov1995}. Mercury, Mars, and the Moon have been in a stagnant-lid mode for all or most of their history; Venus, also lacking evidence for plate tectonics at present, may have instead experienced one or more global resurfacing events in the past, possibly associated with episodes of subduction \citep[e.g.][]{turcotte1993}.

Understanding to what extent a stagnant-lid or plate-tectonic planet may be habitable requires a joint effort
involving the modelling of mantle melting and volcanic outgassing and the characteristics of the resulting
atmosphere. \citet{kite2009} investigated the evolution of melting and volcanism over the entire lifetime of the
parent star for planets with Earth-like structures and compositions, masses between 0.25 and 25 M$_\oplus$, in both
a plate-tectonic and stagnant-lid mode of convection. These authors found that while plate tectonics generally
allows for melt production and outgassing over the entire stellar evolution, for stagnant-lid bodies these
processes tend to cease within a few billion years. Nevertheless, during this time, outgassing rates are predicted
to be significantly higher for stagnant-lid than for plate-tectonic bodies and the planet's mass plays a relatively minor role. \citet{orourke2012} focused specifically on modelling the interior evolution of stagnant-lid planets up to 10 M$_\oplus$ and argued that bodies more massive than the Earth might escape their stagnant-lid regime by producing a significant amount of negatively buoyant crust. Deep crustal layers of super-Earths, in fact, may lie well in the stability field of eclogite; because it is denser than the average peridotitic mantle, eclogite might promote some form of surface mobilization induced by crustal subduction. \citet{noack2014a} used simulations of finite-amplitude convection to investigate the evolution of CO$_2$ outgassing on hypothetical stagnant-lid planets with Earth-like radius but different core sizes. These authors concluded that the steep slope of the melting temperature characterizing the mantle of planets with large cores (i.e. high average density) strongly prevents the generation of partial melt and, in turn, CO$_2$ outgassing, thereby drastically reducing the potential habitability of this kind of terrestrial bodies. 

However, on the basis of interior modelling only, without explicitly coupling the calculated outgassing rates of
greenhouse volatiles with climate simulations, it is difficult to assess precisely the potential for habitability
of a stagnant-lid (or plate-tectonic) body. Atmosphere modelling studies that calculate the boundaries of the HZ
\citep[such as][]{kasting1993a, kopparapu2013, wordsworth2013} usually need to assume a range of atmospheric
pressures and concentrations that are plausible from solar system observations. These studies mainly evaluate the
atmospheric processes that impact the planetary climate and may limit the habitability of the planet. To calculate
the outer edge of the HZ, following \citet{kasting1993a}, it is generally assumed that the planet may provide the appropriate amount of CO$_2$ needed to reach the maximum greenhouse limit via an Earth-like carbon--silicate cycle that stabilizes the planetary climate.

In this work we aim to address the questions of whether and how long a stagnant-lid planet could be habitable, i.e. host liquid water on its surface. To this end, we focus on modelling the coupled evolution of the interior and atmosphere of a hypothetical terrestrial planet with the same mass, radius, and composition as the Earth, but that is characterized by stagnant-lid tectonics throughout its history. We adopt a one-dimensional (1D) model of the thermal evolution of the crust, mantle, and core combined with detailed parameterizations of mantle melting and volatile extraction. The amount of outgassed volatiles (only H$_2$O and CO$_2$ are considered here) then provides the input for our 1D cloud-free radiative-convective atmosphere model that we use to calculate the evolution of the climate of the planet at 1 au from a Sun-like star along with the boundaries of the HZ. In Sect. \ref{sec_theory} we present in detail our modelling framework, including the interior thermal evolution model (\ref{sec_interior}) with the parameterizations adopted to treat mantle melting (\ref{sec_melting}) and volatile outgassing (\ref{sec_outgassing}), the atmospheric model (\ref{sec_atmos}), and the initial conditions and parameters (\ref{sec_param}). Simulation results are presented in Sect. \ref{sec_results}, where we discuss different aspects related to the evolution of the interior (\ref{sec_interiorevol}), the outgassing of volatiles (\ref{sec_outgasevol}), and the resulting atmosphere (\ref{sec_atmosevol}). Discussion and conclusions follow in Sects. \ref{sec_discussion} and \ref{sec_conclusions}.

\section{Theory and models}\label{sec_theory}
%
\subsection{Thermal evolution of the interior}\label{sec_interior}
%
We employ a 1D model of parameterized stagnant-lid convection \citep[e.g.][]{grott2011,morschhauser2011} to simulate the thermal evolution of the interior of an Earth-like planet over 4.5 Gyr starting from a post-accretion scenario when core formation and magma ocean solidification are completed. Although this parameterized approach cannot capture the complexity of the dynamics of the mantle, it compares well with 2D and 3D simulations of the evolution of stagnant-lid bodies such as Mars and Mercury, both in terms of thermal evolution \citep{plesa2015} and crust formation \citep{tosi2013c}.

Given an initial temperature profile for the entire planet (Fig. 1), we solve numerically the time-dependent energy balance equations for the core, mantle, and stagnant lid from which we obtain the evolution of the core-mantle boundary (CMB) temperature, mantle temperature beneath the lid, and thickness of the lid itself, respectively. Energy conservation for the core is given by
\begin{equation}\label{eq_Tc}
    \rho_c c_c V_c \frac{\mathrm{d} T_c}{\mathrm{d} t} = -q_c A_c,
\end{equation}
where $t$ is the time, $\rho_c$, $c_c$, and $V_c$ are the density, heat capacity, and volume of the core, respectively, $T_c$ and $A_c$ are the temperature and surface area of the CMB, and $q_c$ is the heat flux out of the core into the mantle. For simplicity, we neglect in Eq. (\ref{eq_Tc}) the effects of core freezing, i.e.  the accompanying release of latent heat of solidification and gravitational potential energy. 
\begin{figure*}[!ht]
\sidecaption
  \includegraphics[width=12cm]{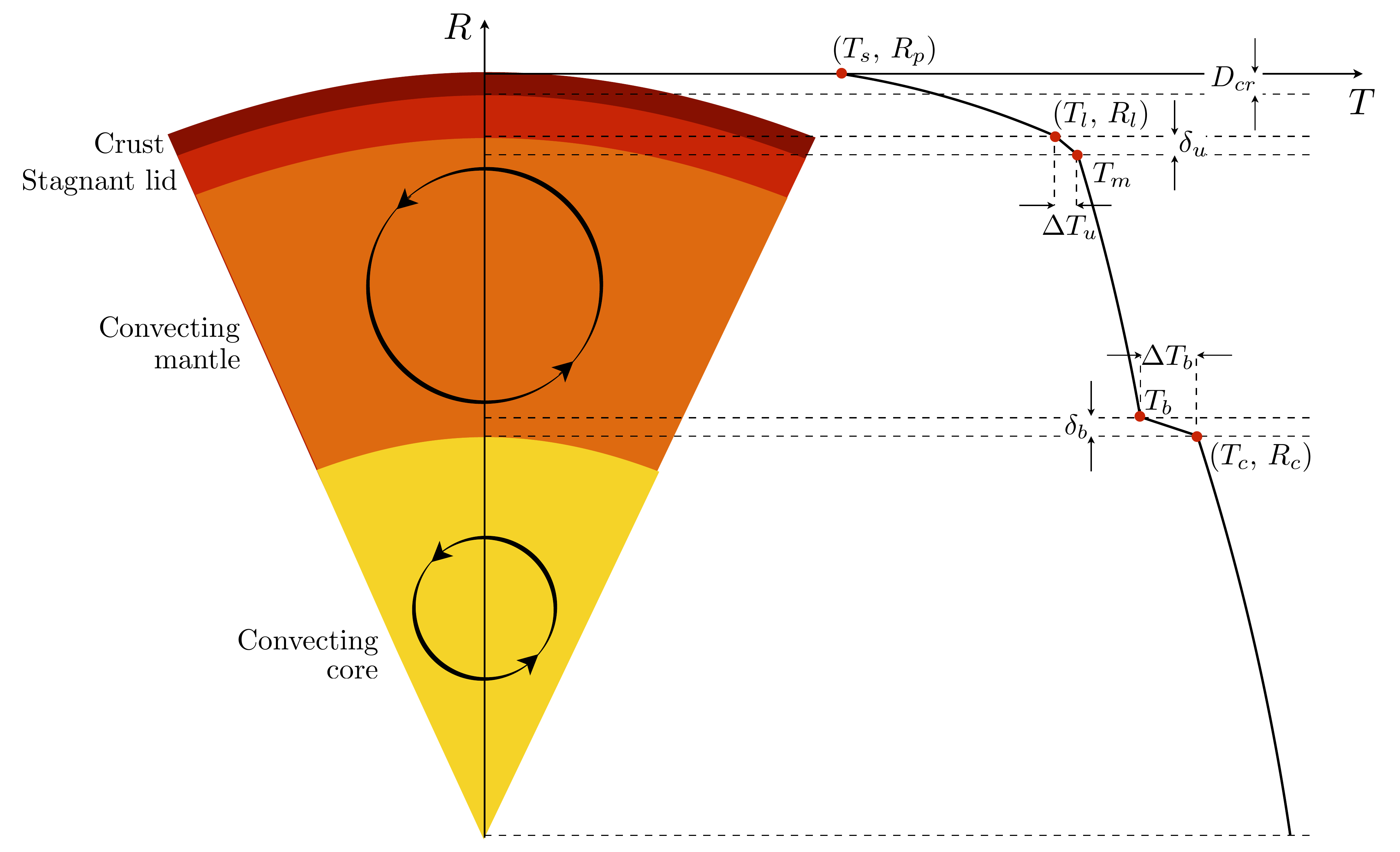}
     \caption{Diagram of the interior structure considered for the thermal evolution model and a schematic of the corresponding temperature profile.
    See text for the description of the various symbols.}
    \label{fig_sketch}
\end{figure*}

The mantle energy balance is governed by the following equation:
\begin{align}\label{eq_Tm}
    & \rho_m c_m V_l (1 + \mathit{St}) \frac{\mathrm{d} T_m}{\mathrm{d} t} = \nonumber \\ & -\left\lbrace q_l + [\rho_{cr} \mathcal{L} + \rho_{cr} c_{cr} (T_m - T_l)]\frac{\mathrm{d}D_{cr}}{\mathrm{d}t} \right\rbrace A_l + q_c A_c + Q_m V_l
\end{align}
%
%
%
where $\rho_m$ and $c_m$ are the density and heat capacity of the mantle; $V_l$ and $A_l$ are the volume and outer area of the convecting part of the mantle, i.e. between the CMB and base of the stagnant lid (see Fig.~\ref{fig_sketch}); $\mathit{St}$ is the Stefan number, which controls the release and consumption of latent heat upon mantle melting and solidification; $T_m$ and $T_l$ are the temperature of the upper mantle and base of the stagnant lid; $\rho_{cr}$, $c_{cr}$, and $D_{cr}$, are the density, heat capacity, and thickness of the crust; $\mathcal{L}$ is the latent heat of melting; $q_l$ and $q_c$ are the (parameterized) heat fluxes from the convecting mantle into the stagnant lid and from the convecting core into the mantle; and $Q_m$ is the mantle volumetric heating rate. On the right-hand side of Eq. (\ref{eq_Tm}), the term proportional to the crust growth rate $\mathrm{d}D_{cr}/\mathrm{d}t$ accounts for the additional heat loss due to the transport of melt from the mantle source region to the surface, i.e. for the so-called heat-piping effect \citep{spohn1991}.

The evolution of the stagnant lid is obtained from the energy balance at its base \citep[e.g.][]{spohn1991}, i.e.
\begin{align}
    & \rho_m c_m (T_m - T_l) \frac{\mathrm{d} D_l}{\mathrm{d}t} = \nonumber \\
    & -q_l + \left[ \rho_{cr} \mathcal{L} + \rho_{cr} c_{cr} (T_m - T_l)\right] \frac{\mathrm{d}D_{cr}}{\mathrm{d}t} - k_m \left. \frac{\partial T}{\partial r}\right|_{r=R_l}, \label{eq_Dl} 
\end{align}
%
%
%
%
where $D_l$ is the thickness of the lid, $k_m$ the thermal conductivity of the mantle, and $\partial T/\partial r$ the radial temperature gradient calculated at the base of the lid (i.e.~at a radius $r=R_l$). The latter is obtained assuming steady-state heat conduction, i.e.
\begin{equation}\label{eq_lidcond}
    \frac{1}{r^2}\frac{\partial}{\partial r}\left(r^2 k_l \frac{\partial T}{\partial r} \right) + Q_l=0,
\end{equation}
where $r$ is the radial coordinate, $k_l$ the thermal conductivity, and $Q_l$ the heat production rate in the stagnant lid. Since this generally comprises both the crust and part of the mantle (Fig.~\ref{fig_sketch}), $k_l$ and $Q_l$ are replaced by the corresponding thermal conductivity and internal heating rate ($k_{cr}$ and $Q_{cr}$ for the crust or $k_m$ and $Q_m$ for the mantle) as appropriate. Although neglecting the time dependence in Eq. (\ref{eq_lidcond}) could affect the earliest transient phases of the evolution, this approximation is sufficiently accurate to capture the long-term thermochemical behaviour of the interior reliably, as demonstrated by comparisons of this approach with the outcomes of fully dynamic simulations \citep{tosi2013c,plesa2015}.

The convective heat fluxes from the core into the mantle ($q_c$) and mantle into the stagnant lid ($q_l$) are obtained from boundary layer theory \citep[e.g.][]{turcotte2002}, which is used to determine the thickness of the two corresponding thermal boundary layers from scaling laws appropriate for stagnant-lid convection \citep{grasset1998}. In particular, the heat flux due to convection in the sublithospheric mantle is proportional to $Ra^{1/3}$, where $Ra$ is the thermal Rayleigh number defined as
\begin{equation}\label{eq_Ra}
    Ra = \frac{\rho_m\alpha g\Delta T (R_l - R_c)^3}{\eta\kappa_m},
\end{equation}
with the coefficient of thermal expansion $\alpha$, the gravitational acceleration $g$, and the mantle thermal diffusivity $\kappa_m=k_m/(\rho_m c_m)$. The superadiabatic temperature difference  $\Delta T$ that drives convection is given by the sum of the temperature drops across the upper and lower thermal boundary layers
\begin{equation}\label{eq_deltaT}
    \Delta T = \Delta T_u + \Delta T_d,
\end{equation}
where $\Delta T_u = T_m - T_l$ is the temperature drop across the top thermal boundary layer and $\Delta T_d = T_c - T_b$ that across the bottom boundary layer ($T_b$ is the mantle temperature right above it). In the definition of Rayleigh number (Eq.~\ref{eq_Ra}), the mantle viscosity $\eta$ is calculated following \citet{hirth2003} assuming an Arrhenius law for wet diffusion creep as follows:
\begin{equation}\label{eq_eta}
    \eta = \frac{A}{X_m^\mathrm{H_2O}} \exp\left(\frac{E^* + P_m V^*}{RT_m}\right), 
\end{equation}
where $A$ is a pre-exponential factor, $R$ is the gas constant, $X_m^\mathrm{H_2O}$ is the water concentration in
the mantle expressed in ppm (see Sect. \ref{sec_melting}), $E^*$ and $V^*$ are activation energy and activation
volume for diffusion creep, and $P_m$ is the pressure at the depth at which the upper mantle temperature $T_m$ is
calculated; see Table \ref{Tab_interior_parameters} for the values of these parameters and Fig.~\ref{fig_Tsol_eta}b
for typical viscosity profiles calculated with Eq.~(\ref{eq_eta}).

\begin{table}[!ht]
\caption{Main parameters of the interior model.}\label{Tab_interior_parameters}
\centering
\resizebox{\columnwidth}{!}{%
\begin{tabular}{l l l}
  \hline
  \hline
  Parameter & Description & Value \\ 
  \hline
  $R_p$ & Planet radius & 6370 km \\
  $R_c$ & Core radius   & 3480 km \\
  $g$   & Gravitational acceleration & 9.8 m s$^{-2}$ \\
  $T_s$ & Surface temperature        & 293 K        \\
  $\Delta T_d$ & Initial core-mantle temperature drop & 200 K\\
  $Q_m$ & Initial mantle heat production & 23 pW kg$^{-1}$ \\
  $\rho_{cr}$  & Crust density  & 2900 kg m$^{-3}$ \\
  $\rho_{m}$   & Mantle density & 3500 kg m$^{-3}$ \\
  $c_{m}$      & Mantle heat capacity & 1100 J kg$^{-1}$ K$^{-3}$ \\
  $c_{c}$      & Core heat capacity   & 800  J kg$^{-1}$ K$^{-3}$ \\ 
  $A$          & Viscosity pre-factor &  $6.127 \times 10^{10}$ Pa s \\  
  $E^*$        & Activation energy    &  $3.35 \times 10^5$ J mol$^{-1}$ \\  
  $V^*$        & Activation volume    &  $4 \times 10^{-6}$ m$^{3}$ mol$^{-1}$ \\
  $k_{cr}$     & Crust thermal conductivity   & 3 W m$^{-1}$ K$^{-1}$ \\
  $k_{m}$      & Mantle thermal conductivity  & 4 W m$^{-1}$ K$^{-1}$ \\
  $\kappa_{m}$ & Mantle thermal diffusivity   & $10^{-6}$ m s$^{-2}$ \\
  $\alpha_{m}$ & Mantle thermal expansivity   & $2 \times 10^{-5}$ K$^{-1}$ \\
  $\mathcal{L}$ & Latent heat of melting      & $6 \times 10^{5}$ J kg$^{-1}$ \\
  $Ra_{crit}$   & Critical Rayleigh number    & 450 \\
  $u_0$         & Convection velocity scale   & $2 \times 10^{-12}$ m s$^{-1}$ \\
  \hline
\end{tabular}
}
\end{table}

In order to calculate the temperature difference in Eq.~(\ref{eq_deltaT}), the temperatures at the base of the lid ($T_l$) and at the base of the mantle ($T_b$) need to be determined.  The latter is readily found by assuming that the mantle is vigorously convecting so that its radial thermal profile is adiabatic and using boundary layer theory to compute the thickness of the bottom thermal boundary layer to the top of which the adiabatic profile extends (see Fig. 1 and Eq. \ref{eq_db}). The lid temperature is obtained instead from scaling laws derived from numerical convection models with strongly temperature-dependent viscosity \citep[e.g.][]{grasset1998,choblet2000}. According to these models, $T_l$ can be identified with the temperature at which the viscosity has grown by about one order of magnitude with respect to the viscosity of the convecting mantle. The lid temperature can then be expressed in terms of the mantle temperature and activation energy as \citep[][]{grasset1998}
\begin{equation}
    T_l = T_m - \Theta\frac{RT_m^2}{E^*},
\end{equation}
with the factor $\Theta$ set to 2.9 to account for the effects of spherical geometry \citep{reese2005}. 

The convective heat fluxes out of the mantle ($q_l$) and core ($q_c$) are
\begin{equation}
    q_l = k_m \frac{T_m - T_l}{d_m}
\end{equation}
and
\begin{equation}
    q_c = k_m \frac{T_c - T_b}{d_b},
\end{equation}
where $d_m$ and $d_b$ are the thicknesses of the upper and lower thermal boundary layers, respectively. According to boundary layer theory, the first is given by \citep[e.g.][]{turcotte2002}
\begin{equation}
    d_m = (R_l - R_c) \left(\frac{Ra_{cr}}{Ra}\right)^{1/3},
\end{equation}
where $Ra_{cr}$ is the critical Rayleigh number for the mantle. The second is given by
\begin{equation}\label{eq_db}
    d_b = \left(\frac{\kappa_m f_c \eta_c Ra_{i,cr}}{\alpha\rho_mg (T_c-T_b)}\right)^{1/3},
\end{equation}
where $f_c$ is a factor accounting for the pressure dependence of the viscosity, $\eta_c = \eta(T_b + T_c)/2$ is the viscosity calculated at the average temperature attained in the lower thermal boundary layer, and $Ra_{i,cr}$ is the local critical Rayleigh number. Following \citet{deschamps2001}, this is given by
\begin{equation}
    Ra_{i,cr} = 0.28 Ra_i^{0.21},
\end{equation}
where $Ra_i$ is the thermal Rayleigh number for the entire mantle, i.e.
\begin{equation}\label{eq_rayleigh}
    Ra_{i} = \frac{\rho_m\alpha g\Delta T_i (R_p - R_c)^3}{\eta\kappa_m},
\end{equation}
with $\Delta T_i = (T_m - T_s) + (T_c - T_b)$.

With the above equations, the thermal evolution of the interior is obtained by advancing in time a radial temperature profile for the entire planet assuming that the temperature increases conductively in the stagnant lid (see Eq.~\ref{eq_lidcond}), linearly in the boundary layers, and adiabatically in the mantle and core. 

It is important to note that the surface temperature ($T_s$) is held constant at 293 K throughout the evolution. Although this may seem inconsistent given that we use an atmospheric model to compute this quantity in response to the time-dependent outgassing of H$_2$O and CO$_2$ (Sect. \ref{sec_atmos}), the effects of taking into account the evolution of $T_s$ are negligible for the interior. Even at the highest surface temperatures obtained from the atmospheric model ($T_s \simeq 430$ K), the temperature-dependence of the viscosity (Eq. \ref{eq_eta}) is sufficiently strong to guarantee that the planet never escapes from a stagnant-lid mode. Indeed, simulations conducted by keeping the surface temperature fixed at 450 K throughout the evolution differ by only 3--5\% in the main output quantities (e.g. average temperature, crustal thickness, and pressure of outgassed volatiles). 

\subsection{Mantle melting, crust production, and element partitioning} \label{sec_melting}
%
The generation of partial melts and the accompanying production of crust are calculated by comparing the mantle temperature profile $T(r)$ with the solidus temperature $T_\text{sol}(r)$, which defines the temperature above which solid rocks begin to melt. At each radius $r$ where the mantle temperature exceeds the solidus, we determine the local melt fraction $\phi(r)$ assuming that it increases linearly between the solidus and liquidus $T_\text{liq}(r)$
\begin{equation}\label{eq_phi}
    \phi(r) = \frac{T(r)-T_\text{sol}(r)}{T_\text{liq}(r) - T_\text{sol}(r)},
\end{equation}
and compute the volume-averaged melt fraction as
\begin{equation}\label{eq_phiave}
    \overline{\phi} = \frac{1}{V_{\phi}}\int_{V_{\phi}}\phi(r)\, \mathrm{d}\phi,
\end{equation}
where $V_{\phi}$ is the volume of the region where partial melting occurs. Since basaltic melts are expected to become denser than the mantle residue at around 8 GPa \citep[e.g.][]{agee2008}, we neglect the extraction of partial melts produced at pressures greater than this value.

\begin{figure}[ht!]
    \centering
    \resizebox{0.7\hsize}{!}{\includegraphics{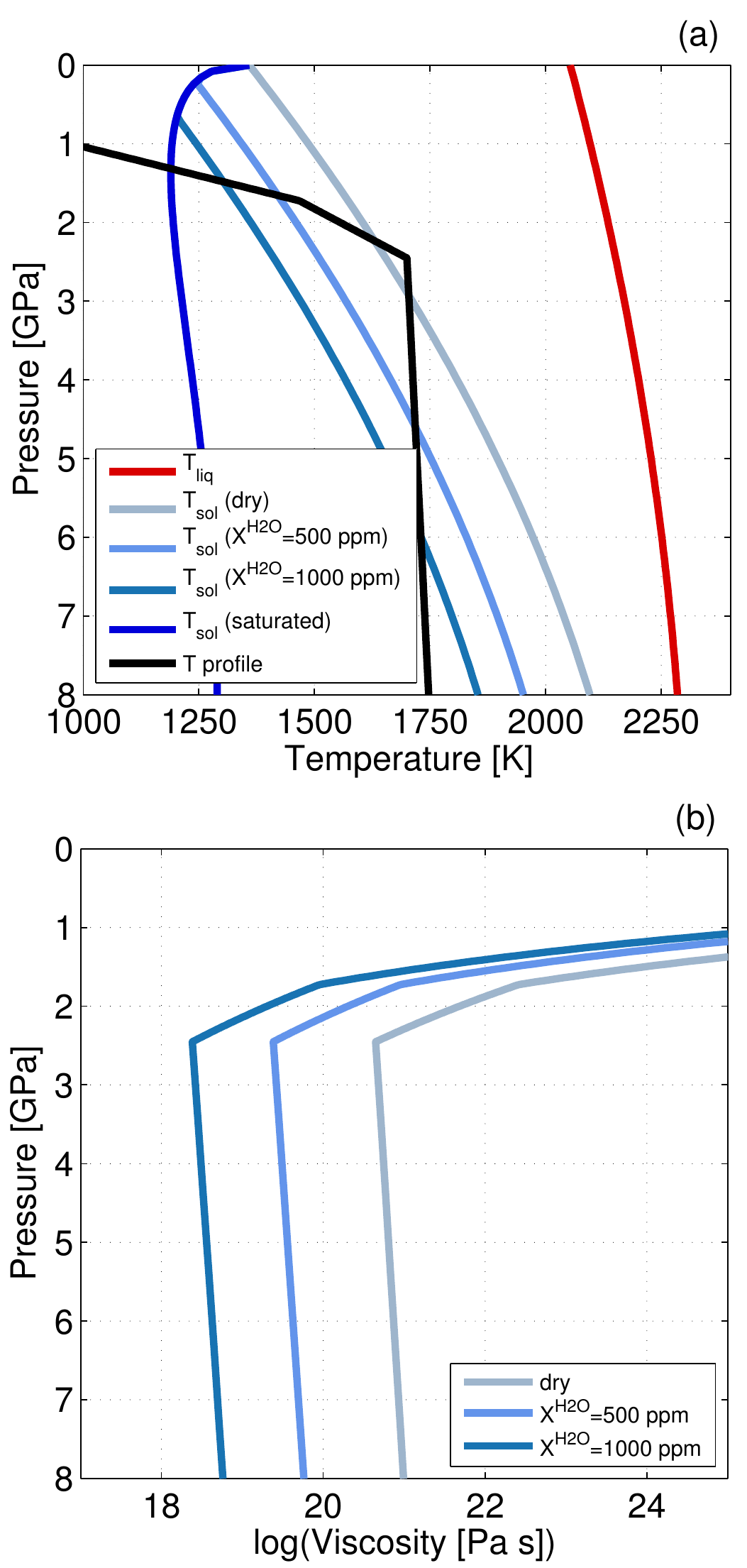}}
    \caption{(a) Solidus temperatures of peridotite for different water contents from dry (grey) to water saturated
    (dark blue), liquidus temperature (red), and a typical initial temperature profile with $T_m=1700$ K (black).
    (b) Viscosity profiles for different water concentrations calculated with Eq. (\ref{eq_eta}) along with the 
    temperature profile shown in panel a.}
    \label{fig_Tsol_eta}
\end{figure}

We consider a peridotitic solidus that we calculate following \citet{katz2003} according to the assumed initial water concentration of the mantle (see Sect. \ref{sec_param}). The presence of water depresses the solidus. Figure \ref{fig_Tsol_eta}a shows solidus profiles for different water concentrations in the mantle from dry (grey line) to water saturated (dark blue line).  For a given value of  water concentration, the actual solidus takes the water-saturated solidus as its lower limit.

Upon melting and subsequent melt extraction, incompatible elements, such as H$_2$O and radiogenic elements, are enriched in the crust and depleted in the mantle. In response to the extraction of water with the melt, the solidus tends to increase. We assume $T_\text{sol}$ to increase linearly with depletion with a maximum change $\Delta T_\text{sol}$ of 150 K corresponding to the solidus difference between peridotite and harzburgite \citep{maaloe2004}.
The timescale for crustal growth depends on the rate at which undepleted mantle material can be supplied to the  partial melt zone, which is a process that takes place according to the mantle convective flow velocity $u$. The crustal growth rate can be thus calculated as
\begin{equation}\label{eq_dcrdt}
    \frac{\mathrm{d}D_{cr}}{\mathrm{d}t} = f_pu\overline{\phi}\frac{V_{\phi}}{4\pi R_p^3},   
\end{equation}
where $f_p$ is a constant that describes the fraction of the surface covered by hot plumes in which partial melting  takes place \citep{grott2010}. The convective velocity is given by
\begin{equation}\label{eq_convvel}
    u = u_0\left(\frac{Ra}{Ra_{crit}}\right)^{2/3},    
\end{equation}
where $u_0$ is the characteristic mantle velocity scale \citep{spohn1991}. The factor $f_p$ in Eq. (\ref{eq_dcrdt}) accounts for the fact that partial melting can either take place in a global sublithospheric channel, where the mantle temperature exceeds the solidus everywhere; this situation corresponds to $f_p=1$ or in the head of isolated mantle plumes covering a limited portion of the surface, in which case $f_p < 1$ \citep{grott2011}. Fully dynamic simulations of large stagnant-lid bodies such as Venus indicate that the planform of mantle convection is likely characterized by a certain number of hot upwellings \citep[e.g.][]{li2007,armann2012,smrekar2012} with partial melting concentrated in plume heads. In the following, we thus consider a plume model with $f_p=0.01$ by adding  a plume excess temperature $\Delta T_d$, corresponding to the temperature drop across the bottom thermal boundary layer, to the mantle temperature used to calculate
melt fractions in Eq. (\ref{eq_phi}).

As we show in Sect. \ref{sec_results}, for most of the evolution our models are characterized by crust forming at a rate that is faster than the rate at which the stagnant lid thickens as a result of mantle cooling.  In this case, we impose that the crust cannot grow thicker than the lid by setting $D_{cr} = D_{l}$, and that, in turn, it is recycled into the mantle by sublithospheric convection, a process that would also be facilitated by the basalt--eclogite transition \citep{orourke2012}.

As mentioned above, we take into account the partitioning of incompatible elements between crust and mantle caused by partial melting. In particular, we consider a model of accumulated fractional melting for the extraction of heat-producing elements (HPEs) and water from the mantle and their enrichment in the crust. The concentration $X_\text{liq}$ of a given trace element in the liquid phase can be obtained from its bulk mantle concentration $X_{m}$ as follows:
\begin{equation} \label{eq_fracmelt}
        X_\text{liq} = \frac{X_{m}}{\phi}\left[1 - (1-\phi)^{1/\delta}\right],
\end{equation}
where $\delta$ is an appropriate partition coefficient. 

\begin{figure}[ht!]
    \centering
    \resizebox{0.75\hsize}{!}{\includegraphics{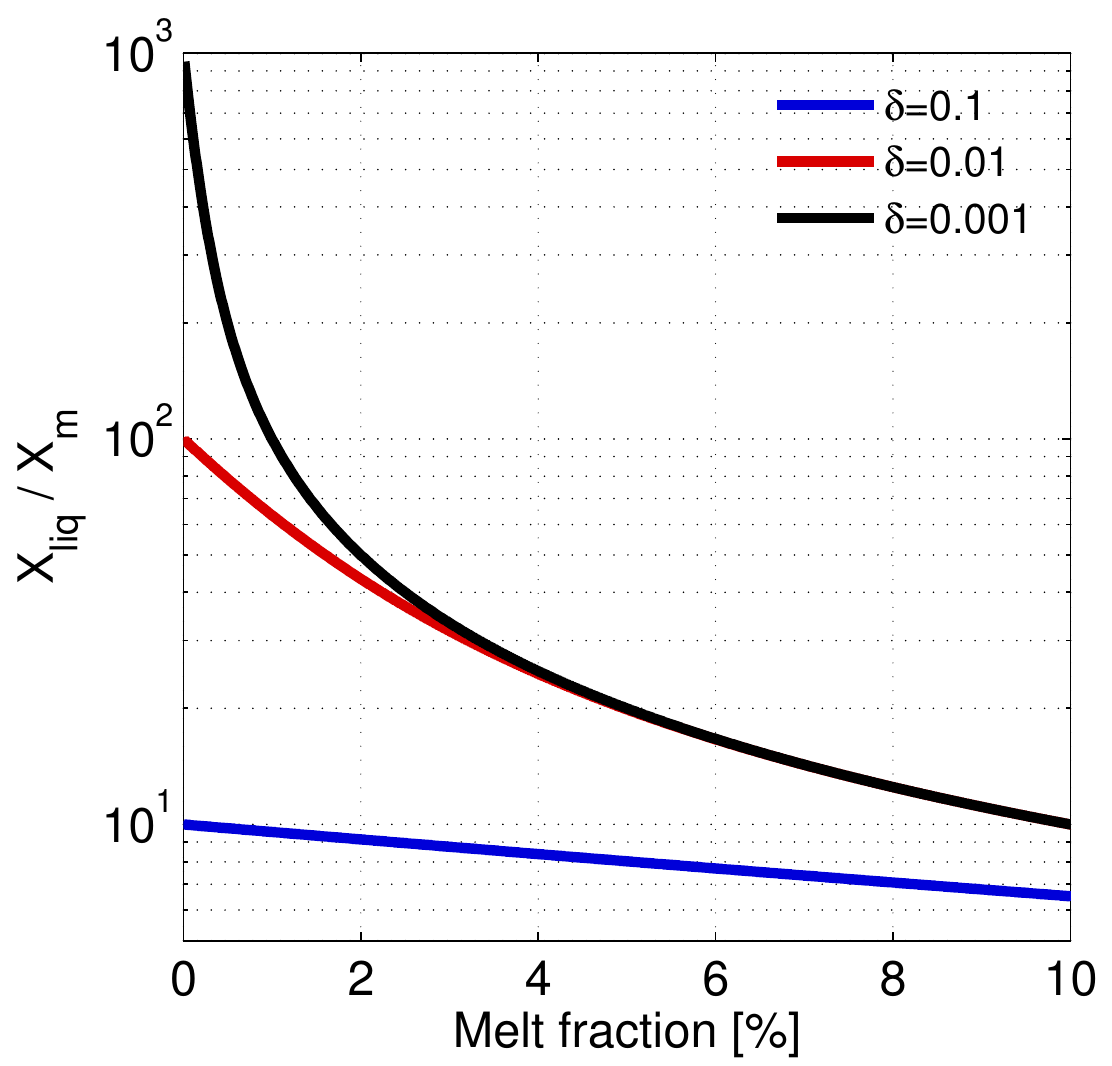}}
    \caption{Ratio of the concentration of incompatible elements enriched in the liquid phase ($X_\text{liq}$) to the corresponding 
            concentration in the solid mantle ($X_{m}$) as a function of melt fraction $\phi$ assuming fractional melting (Eq.
            \ref{eq_fracmelt}) and different partition coefficients $\delta$.}
    \label{fig_fracmelt}
\end{figure}

As shown in Fig. \ref{fig_fracmelt}, for partition coefficients smaller than one, partial melts and, in turn, the crust are strongly enriched in incompatible elements, and even further enriched at small melt fractions. For the long-lived HPEs uranium (U), thorium (Th), and potassium (K), we assumed $\delta=0.001$ \citep[e.g.][]{blundy2003}, while for water, we used $\delta=0.01$ \citep[e.g.][]{aubaud2004}. 

Knowing the depth-dependent melt fraction from Eq. (\ref{eq_phi}), the average concentration of the various incompatible elements in the melt can be readily calculated as
\begin{equation} \label{eq_Xliqave}
    \overline{X}_\text{liq} = \frac{1}{\overline{\phi}V_\phi}\int_{V_\phi} \phi(r)X_\text{liq}\, \mathrm{d}V.
\end{equation}
The total mass of the extracted element ($M_{cr}$) that is enriched in the crust or, in the case of water, partly enriched in the crust and partly outgassed into the atmosphere (Sect. \ref{sec_outgassing}) is proportional to the crust production rate,
\begin{equation} \label{eq_dMcrdt}
        \frac{\mathrm{d}M_{cr}}{\mathrm{d}t} = 4\pi R_p^2\rho_{cr} \overline{X}_\text{liq} \frac{\mathrm{d}D_{cr}}{\mathrm{d}t}.
\end{equation}
Finally, the concentration of the various incompatible elements in the residual mantle is reduced according to the extracted mass,
\begin{equation}\label{eq_Xm}
        X_{m} = \frac{X_{m,0}M_0 - M_{cr}}{M_m},
\end{equation}
where $M_0$ and $M_m$ denote the initial and current mass of the silicate mantle, respectively. Also, Eq. (\ref{eq_Xm}) can be used to calculated the time-dependent volumetric mantle heating rate that is needed in Eq. (\ref{eq_Tm}) as follows:
\begin{equation}
        Q_m(t) = \rho_m\sum_i X_{m,i}(t) H_i \exp(-\lambda_i t),
\end{equation}
where the index $i$ refers to the four long-lived radioactive isotopes $^{235}$U, $^{238}$U, $^{232}$Th, and $^{40}$K, and $\lambda_i$ and $H_i$ are their respective decay constants and specific heat production rates that are chosen according to \citet{mcdonough1995}.  A similar expression holds for the heating rate of the crust,
\begin{equation}
        Q_{cr}(t) = \rho_{cr}\sum_i X_{cr,i}(t) H_i \exp(-\lambda_i t),
\end{equation}
where $X_{cr,i} = M_{cr,i}/M_{cr}$, i.e. the crustal concentration of the various elements is given by the extracted mass divided by the total mass of the crust.  

\subsection{Volatile outgassing}\label{sec_outgassing}

\subsubsection{Outgassing of H$_2$O}\label{sec_h2o-outgas}
As mentioned in the previous section, the extraction of H$_2$O from the interior is determined self-consistently according to a fractional melting model from which, using Eq. (\ref{eq_Xliqave}), we can calculate the average water concentration in the melt. The buoyant melt percolates from the source region through the lithosphere and crust via porous flow or forming dykes and sills, and eventually part of this melt is extruded at the surface. To calculate the actual amount of water that reaches the surface and can be potentially outgassed into the atmosphere, we thus need to assume a certain value for the ratio of intrusive-to-extrusive volcanism ($r_{ie}$). While this parameter affects the thermal history of the interior only marginally, it can have an important impact on the outgassing evolution as it affects the volume of melt available at the surface in a linear way. However, $r_{ie}$ is difficult to constrain as it could vary with time according to the thickness of the lithosphere below which partial melt is generated and would be influenced by the porosity of the upper crust. For simplicity, we assume here $r_{ie}=2.5$ \citep{grott2011}; this value is intermediate between values as low as 1,  typical of some basaltic shields, and 5 or even more,  characteristic of mid-ocean ridges and other volcanic complexes \citep{white2006}. 

\begin{figure}[ht!]
	\centering
	\resizebox{0.8\hsize}{!}{\includegraphics[width=\textwidth]{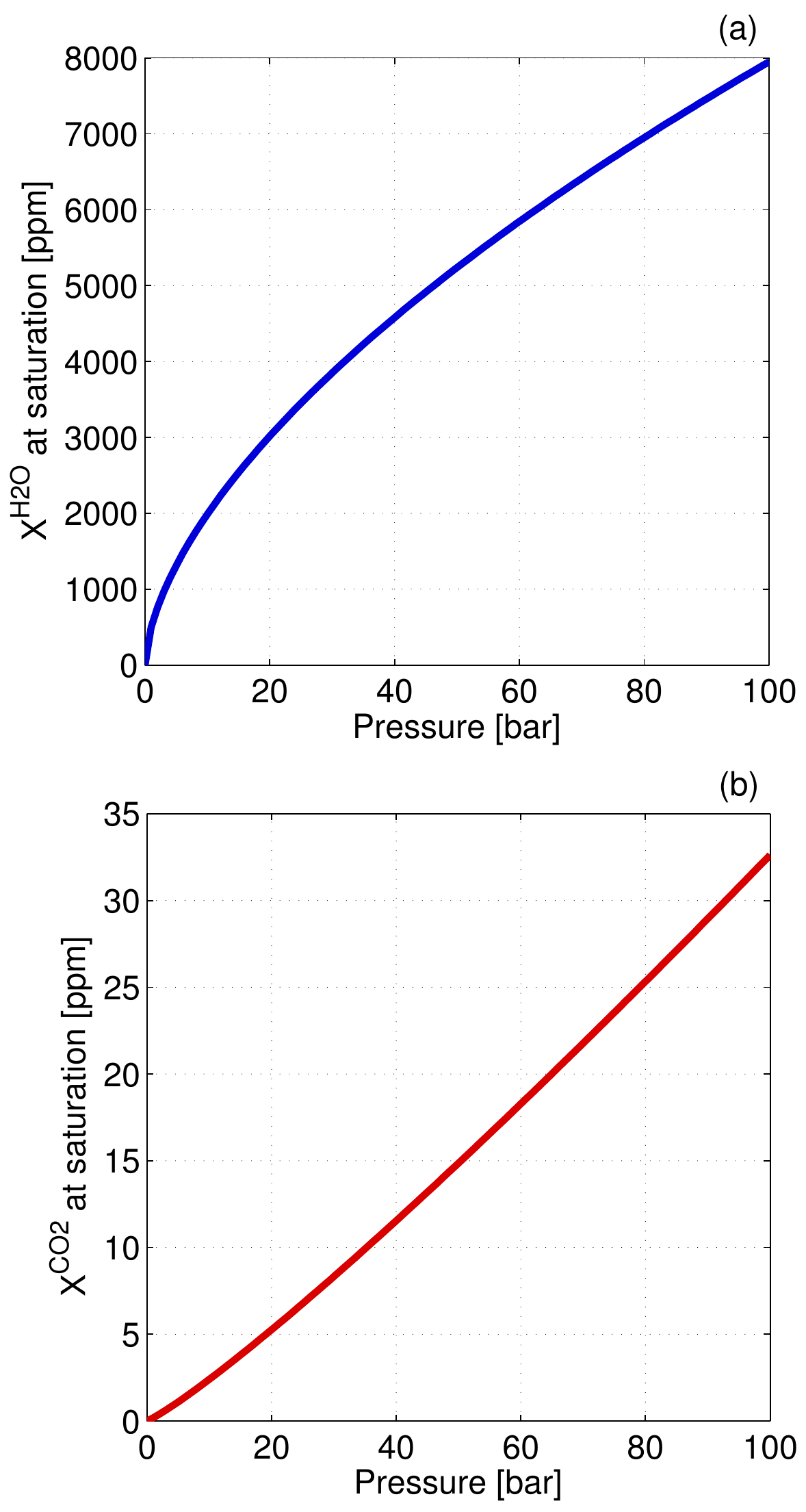}}
    \caption{Saturation concentrations of H$_2$O and CO$_2$ in basaltic melts as a function of pressure. When surface melts are super-saturated, the excess pressure of the corresponding volatile is released into the atmosphere.}
    \label{fig_solubilities}
\end{figure}

Whether the water contained in the extruded melts is outgassed into the atmosphere or retained in the solidifying melt depends on its solubility in surface lavas at the evolving pressure and temperature conditions of the atmosphere \citep{gaillard2014}, whereby the effect of pressure dominates. Figure \ref{fig_solubilities} shows the solubility of H$_2$O (\ref{fig_solubilities}a) and CO$_2$ (\ref{fig_solubilities}b) in basaltic melts as a function of pressure from \citet{newman2002}.  At each timestep of our simulations, we thus check whether the concentration of H$_2$O and CO$_2$ in surface lavas is large enough for these to be supersaturated in the two gases. In this case, the excess concentration is released into the atmosphere, which is then progressively built up, with the partial pressure of water calculated as
\begin{equation}
    P_\mathrm{H_2O} = \frac{M_\text{gas}^\mathrm{H_2O} g}{4\pi R_p^2},
\end{equation}  
where $M_\text{gas}^\mathrm{H_2O}$ is the mass of supersaturated water that can be effectively outgassed.

The solubility of H$_2$O is much larger than that of CO$_2$ (by more than two orders of magnitude at
atmosphere-relevant pressures below 100 bar). As we show in Sect. \ref{sec_results}, it turns out that the
outgassing of water can be significantly limited by its high solubility in the melts, while all extracted CO$_2$ is
easily released into the atmosphere. 

\subsubsection{Outgassing of CO$_2$}\label{sec_co2-outgas}
Modelling of CO$_2$ extraction and outgassing is complicated by the fact that carbon is not directly soluble in silicate minerals, but occurs in separate phases depending on pressure, temperature, and oxygen fugacity \citep[e.g.][]{dasgupta2010}. Under oxidizing conditions, carbon can be present as solid or molten carbonate, while under reducing conditions it occurs in one of its elemental high-pressure forms, i.e. graphite or diamond. Carbonate sediments form as a consequence of the interaction between atmospheric CO$_2$ dissolved in rainwater and silicate rocks. Therefore, in contrast to the Earth where plate tectonics has been operating for billions of years, in stagnant-lid bodies where surface materials cannot be recycled into the interior via subduction, the mantle is likely to be characterized by relatively reducing conditions throughout its evolution. Indeed, there is evidence that the mantle of the Earth was more reduced for much of the Archean than it is now \citep[e.g.][]{AuSt16} and that it has become more and more oxidized in response to hydrogen escape \citep{catling2001} and to the subduction of oxidizing agents such as ferric iron, water, and carbonates \citep{kasting1993b,lecuyer1999}. In contrast, on Mars and the Moon, two bodies that have been likely characterized by a stagnant lid for most or all of their evolution, basalts are generally reduced with oxygen fugacities ($f_\mathrm{O_2}$) ranging from about one log$_{10}$-unit below the iron--w\"ustite buffer (IW) to one unit above it \citep[e.g.][]{herd2002,wadhwa2008}. Nevertheless, in the case of Mars, the composition of basaltic SNC meteorites, from which such reducing conditions are inferred, differs from that of old surface basalts found in the Gusev crater that require a more oxidized source instead \citep{mcsween2009}. An explanation for this discrepancy is that meteorites and surface rocks formed from melting and crystallization of the same source but under different $f_\mathrm{O_2}$ conditions \citep{tuff2013}. The composition of the Gusev crater rocks could be then explained if oxidized surface material was recycled into the upper mantle early in the history of Mars, possibly through an episode of subduction \citep{mcsween2003}. Our stagnant-lid models assume there is no effective mechanism for surface recycling, thus the oxidation state of the mantle should not change significantly over the evolution.

Consistent with this picture, we make the assumption here that the mantle oxygen fugacity is sufficiently low for carbon to be available in its reduced form, which, at the pressure and temperature conditions where partial melting occurs in our models, is likely graphite, or possibly diamond \citep[e.g.][]{stagno2013}. Upon partial melting, some graphite dissolves into the melt in the form of carbonate ions and upon migration and extraction is subsequently outgassed as CO$_2$ at the surface. We follow the approach applied by \citet{grott2011} to model the outgassing of CO$_2$ on Mars, which in turn draws from the thermodynamic framework of redox melting introduced by \citet{hirschmann2008} to assess the solubility of CO$_2$ in graphite-saturated magmas. The abundance of CO$_2$ in the melt ($X_\text{liq}^\mathrm{CO_2}$) depends on the concentration of carbonate ($X_\text{liq}^\mathrm{CO^{2-}_3}$) and on the assumed oxygen fugacity ($f_{\text{O}_2}$) \citep{grott2011} as follows:
\begin{equation} \label{eq_XCO2a}
        X_\text{liq}^\mathrm{CO_2} = \frac{b X_\text{liq}^\mathrm{CO^{2-}_3}}{1 + (b - 1)X_\text{liq}^\mathrm{CO^{2-}_3}},
\end{equation}
where $b$ is a constant appropriate for Hawaiian basalts, and
\begin{equation} \label{eq_XCO2b}
        X_\text{liq}^\mathrm{CO^{2-}_3} = \frac{K_{II} K_I f_{\text{O}_2}}{1 + K_{II} K_I f_{\text{O}_2}},
\end{equation}
where $K_{II}$ and $K_{I}$, also appropriate for Hawaiian basalts, are equilibrium constants governing the reactions 
forming CO$_2$ from graphite and oxygen and carbonate ions from CO$_2$ \citep{holloway1998}.

As for H$_2$O and heat producing elements, the average concentration of the CO$_2$ extracted from the mantle
($X_\text{liq}^\mathrm{CO_2}$) is obtained from Eq. (\ref{eq_Xliqave}) and its mass ($M_{cr}^\mathrm{CO_2}$) by solving Eq.
(\ref{eq_dMcrdt}). The mass of CO$_2$ that is actually outgassed ($M_\text{gas}^\mathrm{CO_2}$) is then calculated by comparing
$X_\text{liq}^\mathrm{CO_2}$ with the saturation concentration of Fig. \ref{fig_solubilities}b and accounting for the
extrusive-to-intrusive ratio of volcanism. Finally, the partial pressure of CO$_2$ delivered to the atmosphere is
given by
\begin{equation}
    P_\mathrm{CO_2} = \frac{M_\text{gas}^\mathrm{CO_2} g}{4\pi R_p^2}.
\end{equation}  
%

\subsection{Surface and atmospheric temperature}\label{sec_atmos}

Taking into account the outgassed greenhouse gases H$_2$O and CO$_2$ from the interior, which build up a secondary atmosphere (assuming the planet lost its primordial atmosphere), we employ a 1D radiative-convective, cloud-free, stationary atmospheric model to calculate the resulting atmospheric temperature, pressure, and water content.

This radiative-convective atmospheric model is based on the atmospheric model of \citet{kasting1984a} and
\citet{kasting1984b}. This original model was further improved by \citet{kasting1988}, \citet{kasting1991}, \citet{kasting1993a}, \citet{pavlov2000}, and \citet{segura2003}. It has been adapted to account for H$_2$O- and CO$_2$-dominated planetary atmospheres over wide temperature and pressure ranges (see \citealp{paris2008,paris2010b}). 

The model uses a variable, logarithmic-equidistant pressure grid based on the hydrostatic equilibrium, taking evaporation and condensation of water at the surface into account. The total atmospheric pressure $P$ is given by $P = P_\text{back} + P_\mathrm{H_2O}$, where $P_\mathrm{H_2O}$ is the partial pressure of H$_2$O and $P_\text{back}$ is the sum of the other partial pressures of molecules present in the atmosphere.

The energy transport that determines the temperature profile in the model is calculated via convection in the troposphere and radiative transfer in the stratosphere  using a time-stepping approach. The convection is described by an adiabatic lapse rate based on \citet{ingersoll1969} accounting for evaporation and condensation of H$_2$O and CO$_2$ \citep[see e.g.][]{kasting1993a,paris2010b}.

The radiative transfer is separated into two distinct frequency regimes: one for the shortwave radiation incident from the star (0.2376--4.545\,{\textmu m} subdivided into 38 spectral intervals) and one for the longwave radiation emitted by the planet (1--500\,{\textmu m} subdivided into 25 spectral intervals).

Only H$_2$O and CO$_2$ are considered as absorbing species in the thermal infrared regime. The spectral fluxes of the thermal radiation are obtained from the spectral intensity via a diffusivity approximation and assuming an isotropic upwelling intensity. The frequency dependence of the radiative transfer equation in this wavelength region is treated with the correlated-$k$ method (e.g. \citealp{mlawer1997}, \citealp{goody1989}). The temperature- and pressure-dependent absorption cross sections used to obtain the $k$-distributions are calculated with the line-by-line radiative transfer code SQuIRRL \citep{schreier2001,schreier2003} using the
Hitemp 1995 database \citep{rothman1995}. The $k$-distributions are calculated for temperatures from 100 to 700$\,$K and for pressures from 10$^{-5}$ to 10$^3\,$bar in nine equidistant logarithmic steps.

In addition to line absorption, collision-induced absorption (CIA) processes in the thermal radiative transfer scheme are also considered. Both self and foreign continua are taken into account. The description of the self and foreign continua of H$_2$O and the foreign continuum of CO$_2$ is based on the approach of \cite{clough1989} (CKD continuum). The implementation of the CKD continuum formulation is adopted from the line-by-line model SQuIRRL. The CO$_2$ self-continuum formulation relies on the approach of \citet{kasting1984b}, which results from weak, pressure-induced transitions of the CO$_2$ molecule near 7\,{\textmu m} and beyond 20\,{\textmu m}. Furthermore, we include the N$_2$-N$_2$ CIA as described in \citet{vonParis2013b}.

The radiative processes considered in the shortwave region include molecular absorption by H$_2$O and CO$_2$ as well as Rayleigh scattering by H$_2$O, CO$_2$, and N$_2$. The numerical scheme is based on \citet{kasting1984b} and \citet{kasting1988} and was improved and described in \citet{paris2010a} and \citet{kitzmann2015}. The equation of radiative transfer is solved by a $\delta$-Eddington two-stream approximation \citep{toon1989}. The frequency dependence of the radiative transfer equation in each spectral band is parameterized by a four-term correlated-$k$ exponential sum in each interval (e.g. \citealp{wiscombe1977}). The absorption cross sections for the gaseous absorption of CO$_2$ in the shortwave radiation are taken from \citet{pavlov2000} using the HITRAN 1992 database \citep{rothman1992}. Absorption coefficients for H$_2$O in the shortwave part of the radiative transfer are updated based on the HITRAN 2008 database \citep{rothman2009}.

The cross section $\sigma_{ray,i}(\lambda)$ for the Rayleigh scattering of a specific species $i$ (N$_2$, CO$_2$, and H$_2$O) for the complete spectral range are described by the approach of \cite{vardavas1984}, which is also used by \cite{paris2010b} and \cite{kopparapu2013} as follows:
\begin{equation}\label{Eq_atmos_rayleigh}
    \sigma_{ray,i}(\lambda)=4.577 \times 10^{-21}\left(\frac{6+3D_i}
            {6-7D_i}\right)\frac{r(\lambda)^2}{\lambda^4}, 
\end{equation}
where $\lambda$ is the wavelength in {\textmu m}, the conversion factor $4.577 \times 10^{-21}$ is taken from \citet{allen1973}, $D_i$ is the depolarization factor, and $r(\lambda)=[10^{-5}\times A_i (1+10^{-3}\times B_i/\lambda^2)]^2$ for CO$_2$ and N$_2$, where $A_i$ and $B_i$ are material parameters for the specific molecule $i$. The values for $D_i$, $A_i$, and $B_i$ for N$_2$ and CO$_2$ are taken from \citet{vardavas1984} and \citet{allen1973}.

For H$_2$O in Eq. (\ref{Eq_atmos_rayleigh}), the depolarization factor $D_\mathrm{H_2O}$ is 0.17 \citep{marshall1990}. The refractivity $r(\lambda)$ of water is determined by $r(\lambda)=0.85r_{dry}(\lambda)$ \citep{edlen1966} with the refractivity of dry air $r_{dry}$ approximated by \citet{bucholtz1995},
\begin{equation}
    r_{dry}(\lambda)=\frac{5.7918\times 10^{-2}}{238 -
    \lambda^{-2}}+\frac{1.679\times 10^{-3}}{57.362 - \lambda^{-2}}
.\end{equation}
%

\subsection{Initial conditions and model parameters}\label{sec_param}
%
We ran a series of simulations of the evolution of the interior by varying three main parameters that exert a first-order influence on the outgassing history of the planet. In particular, we varied the initial mantle temperature ($T_{m,0}$) between 1600 and 1800 K, the initial water concentration of the mantle ($X_{m,0}^\mathrm{H_2O}$) between 0 (corresponding to a dry mantle) and 2000 ppm, and the mantle oxygen fugacity ($f_{\text{O}_2}$) between one log$_{10}$-unit below the IW buffer (IW-1) and two log$_{10}$-units above it (IW+2).

The range of temperatures is chosen to limit the initial melt fractions to values below $\sim\! 40$\%, above which the mantle would no longer deform via viscous creep, but rather exhibit a fluid-like behaviour \citep[e.g.][]{costa2009} that the scaling laws we employ to model heat transfer via solid-state convection could not capture. The black line in Fig. \ref{fig_Tsol_eta}a shows the initial temperature profile of the uppermost part of the mantle down to 8 GPa for $T_{m,0} = 1700$ K. Figure \ref{fig_Tsol_eta}b instead shows three initial viscosity profiles calculated with Eq. (\ref{eq_eta}) along the temperature distribution of Fig.  \ref{fig_Tsol_eta}a for a dry mantle and for initial water concentrations of 500 and 1000 ppm.

The present-day upper mantle of the Earth is largely depleted in incompatible elements and, based on studies of mid-ocean ridge basalts, also relatively dry with a water concentration of 50--200 ppm \citep[e.g.][]{saal2002}. Estimates of the bulk Earth water abundance indicate a value within a relatively broad range between 550 and 1900 ppm \citep{jambon1990}. Furthermore, planetary formation models predict that a very large amount of water (up to several Earth oceans) can be stored during the accretion of terrestrial planets that form near 1 au \citep{raymond2004}. Indeed, the H$_2$O storage capacity of nominally anhydrous minerals (olivine and pyroxene) can be as large as few thousand ppm at upper mantle pressures \citep{hirschmann2005}.  By choosing a range between 0 and 2000 ppm, we thus cover a broad parameter space of plausible bulk water contents. 

The oxygen fugacity of meteoritic materials is generally low. For ordinary chondrites, for example, it lies about
three log$_{10}$-units below the IW buffer (IW-3). Shergottites, petrologically the most primitive of the Martian
meteorites, instead have a slightly higher oxygen fugacity between IW and IW+1 \citep{righter2006}. By assuming
$f_\mathrm{O_2}$ to vary between IW-1 and IW+2, we thus cover a broad range of redox conditions, from highly to
moderately reducing, which leads to a similarly broad range of CO$_2$ outgassing rates (see Sect. \ref{sec_outgasevol}).

In all simulations we assumed that the core is initially super-heated with respect to the mantle by 200 K. The impact of this choice however is not very significant since for internally heated stagnant-lid bodies initial differences in the temperature drop across the bottom thermal boundary are rapidly eliminated by efficient extraction of heat from the core \citep[e.g.][]{plesa2015}. 

The atmospheric model is used to calculate snapshots taking into account the outgassing of the greenhouse gases H$_2$O and CO$_2$ from the interior model and additionally the evolution of the luminosity of the Sun \citep[see][]{gough1981}. Atmospheric snapshots are calculated in steps of 0.1\,Gyr from 0.1 to 0.5\,Gyr and in steps of 0.5\,Gyr from 0.5 to 4.5\,Gyr.  The model considers molecular nitrogen (N$_2$), CO$_2$ and H$_2$O as atmospheric gases. These are key component gases in the atmospheres of the terrestrial planets in our solar system. The concentration profile for N$_2$ is an isoprofile of 1$\,$bar. The concentration of CO$_2$, which is also handled as an isoprofile in the model, results from the partial pressure of CO$_2$ ($P_\mathrm{CO_2}$) from the interior model divided by the background pressure $P_\text{back}$. The atmospheric profile of H$_2$O is determined by assuming a completely saturated atmosphere (i.e. a relative humidity of 100\%), such that the partial pressure of H$_2$O ($P_\mathrm{H_2O}$) is determined by the (temperature-dependent) saturation vapor pressure. Therefore, this approach is limited to temperatures below the critical point of water (647 K). The water reservoir is limited by $P_\mathrm{H_2O}$ outgassed from the interior. We assume an amount of N$_2$ of 1 bar that is similar to the reservoirs found on Earth and Venus and, in addition, facilitates the comparison with other habitability studies, for example, by \citet{kasting1993a}.

We account for the solar evolution by increasing the luminosity of the Sun with time $L(t)$ following \citet{gough1981},
\begin{equation}\label{Eq_gough}
    L(t)=L_0\left[1+\frac{2}{5}\left(1-\frac{t}{t_0}\right)\right]^{-1}
\end{equation}
when $t \lesssim t_0$. The value $L_0$ is the present solar luminosity and $t_0$ is the main-sequence lifetime of the Sun, which is 4.7 Ga. Additionally we vary the solar flux to determine the HZ boundaries. The solar input spectrum is based on a high-resolution spectrum of the Sun by \cite{gueymard2004}. The mean solar zenith angle of 60$^{\circ}$ is used in the calculations.

The planetary gravity acceleration is assumed to be the same as for Earth. Furthermore, a surface albedo of 0.22 is assumed, which is larger than the mean observed value of present Earth (0.13). Following the approach of \citet{kasting1988} and \citet{segura2003}, this high value is used to mimic the reflectivity of clouds in the planetary atmosphere. When assuming a relative humidity profile comparable to that of the Earth \citep{manabe1967}, this albedo leads to the global mean temperature of the Earth (288\,K). Assumptions about the relative humidity and surface albedo can have a large impact on the planetary climate as shown, for example, by \citet{Godolt2016}. We chose a relative humidity similar to those in \citet{kasting1993a} and \citet{kopparapu2013}. This allows for a better comparison to these previous studies.

The orbital distance $d$ of the planet to the star is calculated as
\begin{equation}\label{Eq_S_in_d}
    d = 1\,\mathrm{au} \sqrt{\frac{S_0}{S}},
\end{equation}
where $S$ is the solar flux used in the model and $S_0$ the solar constant (i.e. the solar flux at Earth's orbit).

The important parameters used in the following computations are summarized in Table \ref{Tab_summary_parameters}.

\begin{table}[!ht]
\caption{Parameters of the atmospheric model.}
\label{Tab_summary_parameters}
\centering
\resizebox{\columnwidth}{!}{%
\begin{tabular}{l l}
  \hline
  \hline
  Property & Value \\ \hline 
  Stellar spectrum & present Sun                      \\ 
  Solar constant ($S_0$) & 1366$\,$Wm$^{-2}$\\ 
  Solar zenith angle & 60$^{\circ}$ \\ 
  Gravitational acceleration  & 9.8$\,$m\,s$^{-2}$\\ 
  Background surface pressure ($P_\text{back}$) & 1$\,$bar\\ 
  Atmospheric pressure at the top of the atmosphere
  & 6.6$\times10^{-5}\,$bar\\ 
  Atmospheric composition & N$_2$, CO$_2$, H$_2$O \\ 
  Relative humidity & 100\%\\ 
  Surface albedo & 0.22 \\
  \hline
\end{tabular}
}
\end{table}

\section{Results}\label{sec_results}

\subsection{Thermochemical evolution of the interior}\label{sec_interiorevol}

\begin{figure*}[ht!]
	\centering
	\resizebox{0.85\hsize}{!}{\includegraphics{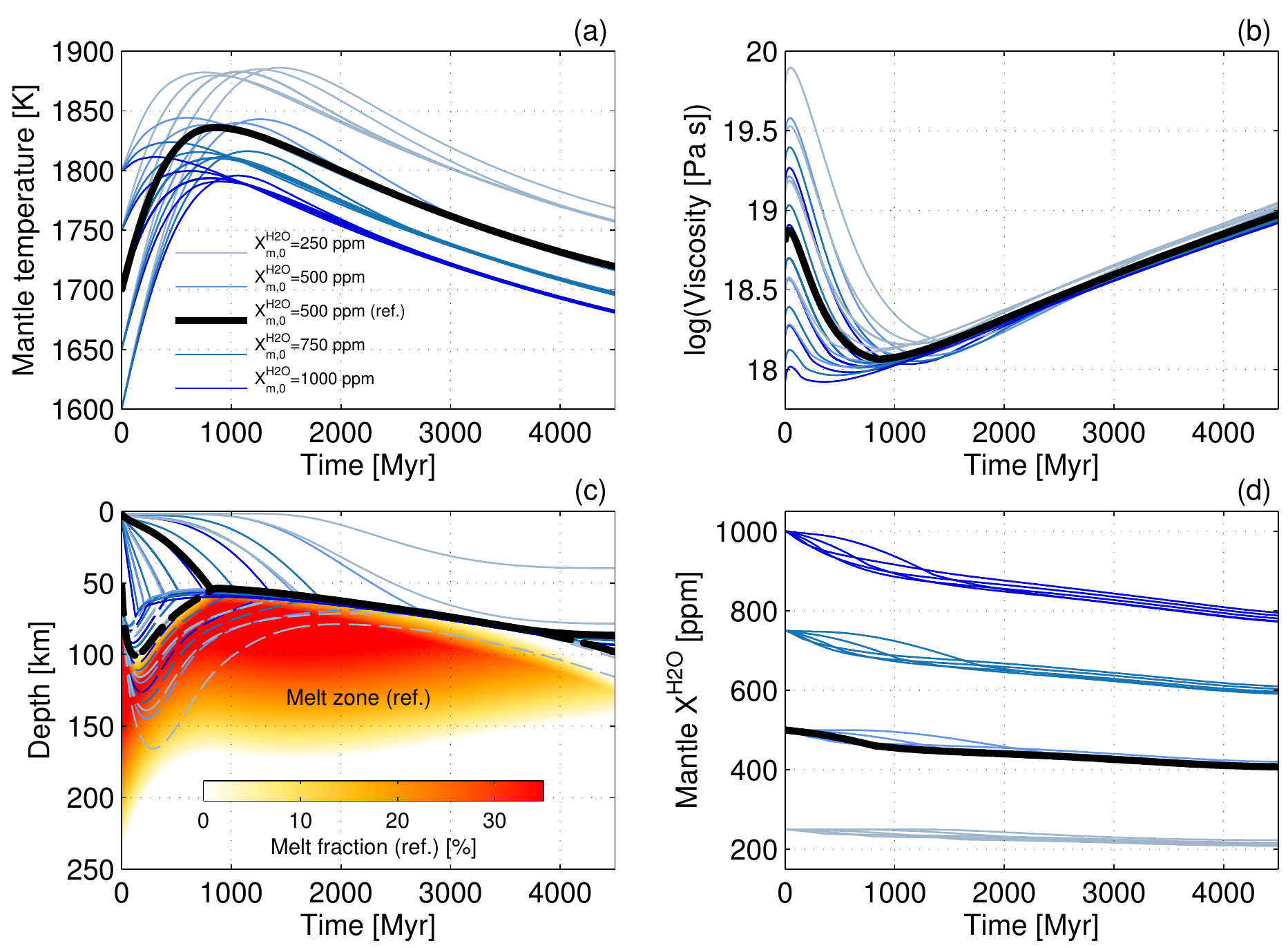}}
    \caption{Evolution of the interior for different initial mantle temperatures between 1600 and 1800 K, initial water concentrations ($X^\mathrm{H_2O}_{m,0}$) between 250 and 1000 ppm, and an oxygen fugacity corresponding to the IW buffer. (a) Mantle temperature is shown; (b) corresponding mantle viscosity is shown; (c) thickness of the crust (solid lines), the stagnant lid (dashed lines), and distribution of the melt zone and melt fraction (coloured area) are shown; and (d) water concentration in the mantle is shown. The different colours, from grey to dark blue, indicate increasing initial water concentrations; the thick black lines in the four panels and the coloured melt zone in panel c indicate the reference model (ref.) with $T_{m,0}=1700$ K, $X^\mathrm{H_2O}_{m,0}=500$ ppm, and $f_\mathrm{O_2}$ at the IW buffer. The different initial mantle temperatures, which can be identified in panel a, are not indicated with distinct colours.}
    \label{fig_interiorevol}
\end{figure*}

The thermochemical evolution of the mantle is summarized in Fig. \ref{fig_interiorevol}, where we show the evolution of the mantle temperature (\ref{fig_interiorevol}a), corresponding viscosity (\ref{fig_interiorevol}b), crustal thickness, depth extent of the melt zone and melt fraction (\ref{fig_interiorevol}c), and concentration of water in the mantle (\ref{fig_interiorevol}d) for a series of simulations with different initial water concentrations and temperatures. The various colours refer to initial H$_2$O concentrations from 250 ppm (grey) to 1000 ppm (dark blue), while no colour distinction is used to indicate the various initial temperatures. The thick black line in the four panels describes the evolution of a reference model characterized by intermediate values of the water content ($X^\mathrm{H_2O}_{m,0}=500$ ppm) and initial mantle temperature ($T_{m,0}=1700$ K). In all models, the mantle oxygen fugacity is set to the IW buffer. The latter, however, albeit fundamental for the outgassing of both CO$_2$ and H$_2$O, has only a secondary effect on the evolution of the interior (see Sect. \ref{sec_outgasevol}). 

As shown in Fig. \ref{fig_interiorevol}a, the thermal history is generally characterized by an initial heating phase during which convective cooling is not efficient enough to remove the internal heat generated by the decay of radioactive elements, a behaviour that is characteristic of the early evolution of the interior of stagnant-lid bodies \citep[e.g.][]{morschhauser2011,tosi2013c}. After this phase, which lasts between 500 and 1500 Myr depending on the model parameters, the mantle and the core (not shown) cool at a roughly constant rate of $\sim\! 40$ K/Gyr. The thermal history is largely controlled by the choice of the initial water concentration of the mantle; the higher is the latter, the lower the mantle viscosity (see Eq. \ref{eq_eta} and Fig. \ref{fig_interiorevol}b). A low viscosity causes convection, and hence heat loss, to be more efficient (Eq. \ref{eq_rayleigh}) with the consequence that models that have the highest initial water content -- and hence the lowest reference viscosity -- are also characterized by the shortest heating phase and, in turn, by the lowest temperature at the end of the evolution (see curves with $X^\mathrm{H_2O}_{m,0}=1000$ ppm in Fig. \ref{fig_interiorevol}a). The effect of the initial mantle temperature on the overall thermal evolution of the interior is minor compared to that of water. Because of the strong exponential dependence of the viscosity on temperature, in fact, an increase in the mantle temperature is accompanied by a viscosity reduction (Eq. \ref{eq_eta}) that promotes convection and leads to a more rapid heat loss. On the contrary, upon cooling, the viscosity increases, slows convection down, and renders heat transfer less efficient. Relatively small changes in the initial temperature produce thus large variations in the heat flux. As a consequence, for a given choice of $X^\mathrm{H_2O}_{m,0}$, the temperature is buffered at a nearly constant value, as expected according to the so-called Tozer effect \citep{tozer1967}. Models with different initial temperatures tend thus to converge rapidly and evolve in a similar fashion. Furthermore, the higher  $X^\mathrm{H_2O}_{m,0}$ is, the earlier differences in the initial temperature tend to be removed. 

Since the solidus temperature strongly depends on the hydration state of the mantle (Fig. \ref{fig_Tsol_eta}a), the initial H$_2$O concentration also has a fundamental influence on the production of partial melt and on the formation of crust (and, in turn, on the outgassing history as we discuss in Sect. \ref{sec_outgasevol}). As shown in Fig. \ref{fig_interiorevol}c, the initial phase of mantle heating leads to the production of a large volume of partial melt, which causes the crust to grow the more rapidly the higher $X^\mathrm{H_2O}_{m,0}$ is. With the exception of some models with low $T_{m,0}$ and $X^\mathrm{H_2O}_{m,0}$, in all cases the crust (solid lines in Fig. \ref{fig_interiorevol}c) stops growing when it becomes as thick as the stagnant lid; there is a sharp bend in the crust curves when they cross the stagnant-lid curves indicated with dashed lines. When this happens, the erosion of the bottom part of the crust starts \citep{morschhauser2011} and continues until the rate at which the stagnant lid thickens because of mantle cooling overcomes the rate at which crust is produced. In our reference model, this phase lasts between $\sim\! 800$ and 3300 Myr (thick black lines in Fig. \ref{fig_interiorevol}c). Nevertheless, melt and crust production continue over the entire evolution of the mantle, albeit at a decreasing rate. Neglecting  the effects of the possible nucleation of an inner core could  affect, at least in principle, the evolution of melt generation. The release of latent heat and of heat associated with the change in gravitational potential energy upon core freezing could  slow down the cooling of the core with respect to the cooling of the mantle. The accompanying increase in the temperature drop across the bottom thermal boundary layer could thus favour the formation of plumes and the production of partial melt, which could potentially lengthen the degassing lifetime of the planet. Although the difference in the cooling rates of mantle and core caused by inner-core freezing is likely to be small \citep{grott2011b}, this effect should be carefully quantified in the future.

Upon melting, incompatible components are removed from the mantle and extracted into the crust or outgassed at the surface. Figure \ref{fig_interiorevol}d shows the evolution of the mantle water concentration. As expected for a mantle undergoing partial melting over its entire evolution, the concentration of water decreases continuously over time, reaching about 80\% of its initial value after 4.5 Gyr.  Similar to the crustal growth rate, the rate of water extraction diminishes at the time when subcrustal erosion starts as a consequence of the recycling of wet crust into the mantle; see e.g. the change in slope of the curve corresponding to the reference model at $\sim\! 800$ Myr in Fig. \ref{fig_interiorevol}d. 

\subsection{Outgassing evolution} \label{sec_outgasevol}

\begin{figure*}[ht!]
	\centering
	\resizebox{0.85\hsize}{!}{\includegraphics{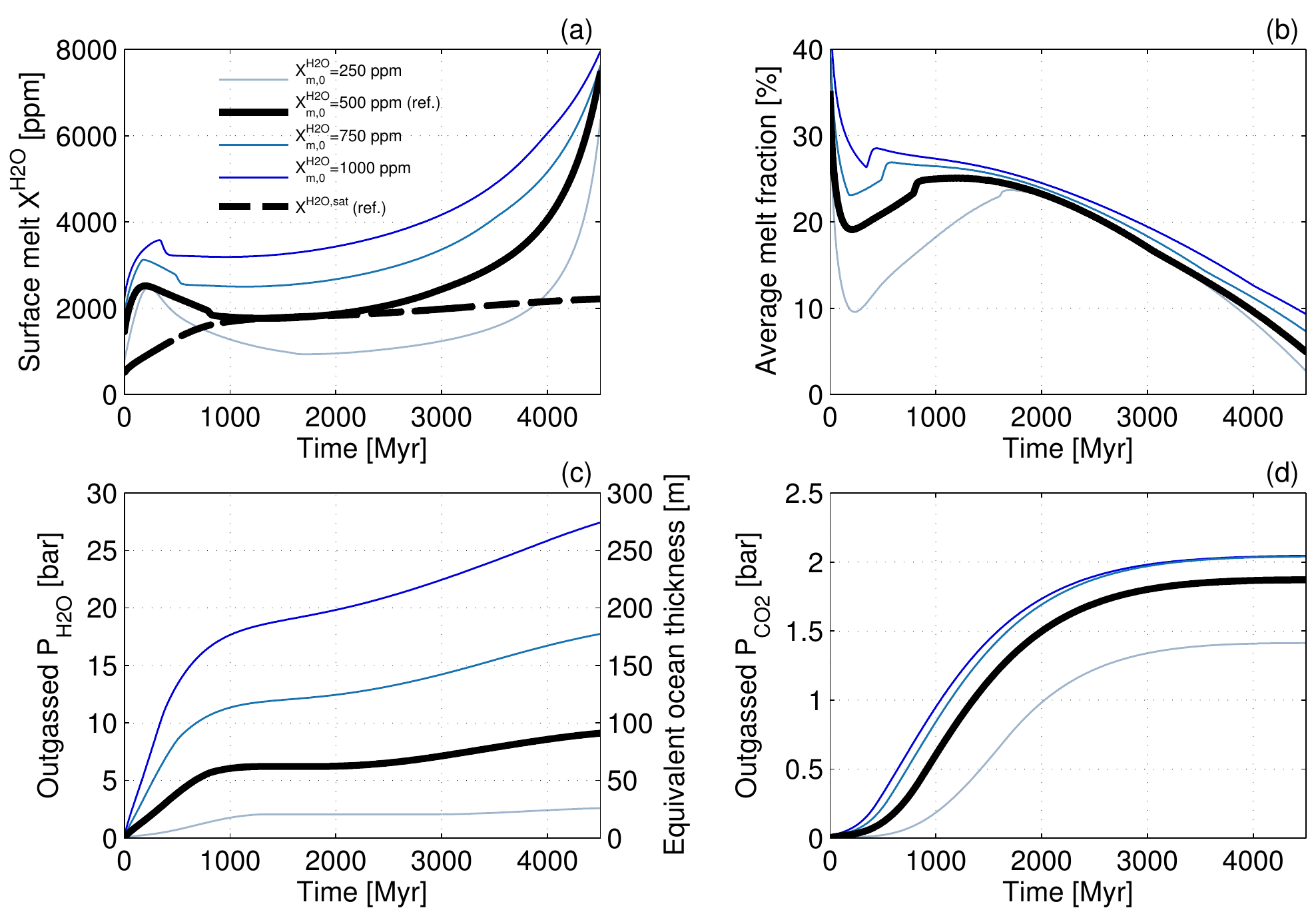}}
    \caption{Extraction and outgassing evolution of H$_2$O and CO$_2$ for models with initial water concentrations ($X^\mathrm{H_2O}_{m,0}$) between 250 and 1000 ppm, an initial mantle temperature of 1700 K, and an oxygen fugacity at the IW buffer. (a) Concentration of H$_2$O in the extracted melt (solid lines) and, for the reference model only, concentration of water at saturation (dashed line) are shown; (b) average melt fraction is shown; (c) partial pressure of outgassed H$_2$O (also expressed as equivalent ocean thickness) is shown; (d) partial pressure of outgassed CO$_2$ is shown.} \label{fig_outgasevol}
\end{figure*}

For a subset of the models discussed in the previous section (only those with $T_{m,0}=1700$ K), Fig. \ref{fig_outgasevol} shows the main features of the extraction and outgassing evolution of H$_2$O and CO$_2$. The possibility that the two volatiles are released into the atmosphere depends on whether their concentration in the melts that are extruded at the surface is higher than their saturation concentration at the evolving pressure conditions of the atmosphere (Sect. \ref{sec_h2o-outgas} and \ref{sec_co2-outgas}). Figure \ref{fig_outgasevol}a shows the evolution of the concentration of H$_2$O in surface melts (solid lines) and the evolution of the saturation concentration (black dashed line for the reference model only). The latter is greater than zero already at the beginning of the evolution because of the background pressure of 1 bar N$_2$ used in the atmospheric model (see Sect. \ref{sec_atmos}).

Since the extraction of water is calculated according to a fractional melting model, the evolution of its surface concentration can be readily explained in terms of the evolution of the average melt fraction shown in Fig. \ref{fig_outgasevol}b. In general, as the degree of partial melting increases because of mantle heating, the concentration of water that partitions into the melt decreases, and vice versa (Eq. \ref{eq_fracmelt} and Fig. \ref{fig_fracmelt}). After a short initial transient phase during which the average melt fraction decreases, it rises over a time interval whose duration is controlled by the mantle viscosity, or, indirectly, by the assumed initial water concentration: the higher $X^\mathrm{H_2O}_{m,0}$ (i.e. the lower the mantle viscosity) is, the shorter is this interval (Fig. \ref{fig_outgasevol}b). The water concentration in the surface melts evolves accordingly; after the initial transient phase, this water concentration first decreases and then rises as the mantle cools and melt fractions decline. Outgassing of H$_2$O can only take place when the water concentration in the surface melts is larger than the saturation concentration, which increases monotonically with time according to the amount of volatiles that are progressively released into the atmosphere. In our reference model, water can be outgassed during the first billion years of evolution and later than about 2.5 Gyr (compare solid and dashed black lines in Fig. \ref{fig_outgasevol}a). As a consequence, the partial pressure of H$_2$O rises to $\sim\! 6$ bar during the first outgassing phase; this pressure remains constant until about 2.5 Gyr while surface melts are undersaturated in water and then rises again to reach $\sim\!9$ bar at the end of the evolution, which would correspond to a global water layer of $\sim\!90$ m (black line in Fig. \ref{fig_outgasevol}c). Despite the high values of water concentration in the surface melt that are achieved during the second outgassing phase, the increase in the partial pressure of atmospheric water is relatively small compared to the first phase because the overall volume of partial melt produced declines significantly with time (Fig. \ref{fig_interiorevol}c) and because the continuous increase of the total atmospheric pressure makes further degassing, of H$_2$O in particular, increasingly difficult.

The outgassing evolution of CO$_2$ is somewhat simpler than that of H$_2$O for two reasons. On the one hand, CO$_2$ outgassing is not calculated on the basis of fractional melting but depends on the assumed oxygen fugacity of the mantle (Sect. \ref{sec_co2-outgas}). On the other hand, since the saturation concentration of CO$_2$ in surface melts is much lower than that of H$_2$O (see Fig. \ref{fig_solubilities}), CO$_2$ tends to be outgassed much more easily. Indeed, its surface concentration (which is not shown in Fig. \ref{fig_outgasevol}) remains above the saturation level of the melt throughout the evolution. As a consequence, the partial pressure of outgassed CO$_2$ rises monotonically as long as partial melt is produced, which, in these models, occurs over the entire evolution. In our reference model, in which we assumed an oxygen fugacity at the IW buffer, it reaches $\sim\!1.8$ bar after 4.5 Gyr (Fig. \ref{fig_outgasevol}d), although the increase in pressure after $\sim\!3$ Gyr is minor owing to the small amount of partial melt produced during the last part of the evolution.

The effect of different initial water concentrations is significant for the outgassing history of both H$_2$O and CO$_2$. As expected, the higher $X^\mathrm{H_2O}_{m,0}$ is, the higher are the final partial pressures of the two gases. In the case of CO$_2$, this is because a high water concentration in the mantle causes a strong decrease of the solidus temperature, which facilitates the production of large volumes of partial melt. In the case of H$_2$O, in addition to the above reason, a high initial concentration clearly makes the amount of water available for outgassing high as well with the consequence that the final partial pressure of outgassed H$_2$O ranges from 2.5 bar for $X^\mathrm{H_2O}_{m,0}=250$ ppm to 27 bar for $X^\mathrm{H_2O}_{m,0}=1000$ ppm.

\begin{figure}[ht!]
	\centering
	\resizebox{\hsize}{!}{\includegraphics{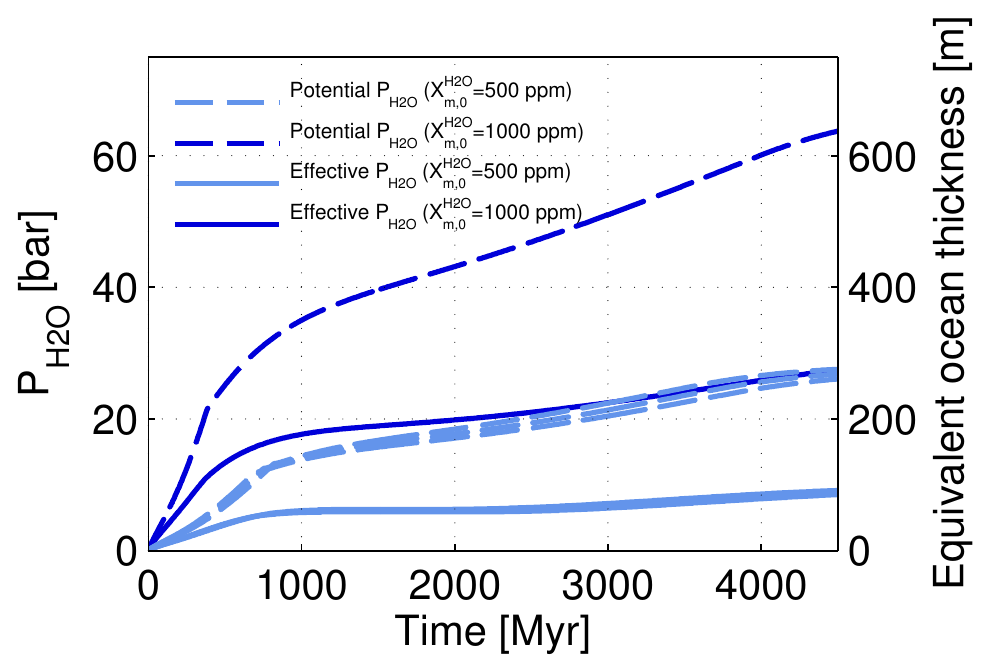}} 
    \caption{Evolution of the partial pressure of outgassed H$_2$O for a model with $T_{m,0}=1700$ K, $f_\mathrm{O_2}$ at the IW buffer, and two initial water concentrations ($X^\mathrm{H_2O}_{m,0}=500$ and 1000 ppm) calculated by neglecting the evolving saturation concentration (dashed lines) and taking such a concentration into account (solid lines). } \label{fig_p-effect}
\end{figure}

As already discussed above and shown in Fig. \ref{fig_outgasevol}c, the outgassing history of H$_2$O is strongly influenced by the evolution of the concentration of water in surface melts. In general, we observed an initial phase during which these are largely supersaturated and H$_2$O is efficiently outgassed. The duration of this phase is inversely proportional to the initial water concentration and varies between 800 and 1200 Myr for $X^\mathrm{H_2O}_{m,0}$ ranging from 1000 to 250 ppm. Afterwards, because of the increase of the saturation concentration from the increasing atmospheric pressure caused by the accumulation of both H$_2$O and CO$_2$, H$_2$O outgassing slows down (for $X^\mathrm{H_2O}_{m,0}=750$ and 1000 ppm) or stops entirely (for $X^\mathrm{H_2O}_{m,0}=250$ and 500 ppm) for a period whose duration is also inversely proportional to $X^\mathrm{H_2O}_{m,0}$. In Fig. \ref{fig_p-effect} we show how significant the effect of evolving saturation conditions at the surface can be for the outgassing history of water. For two initial water concentrations of 500 and 1000 ppm, the figure illustrates the evolution of the partial pressure of H$_2$O outgassed into the atmosphere obtained when the effect of the evolving saturation concentration is neglected (dashed lines) and taken into account (solid lines). In the latter case, the amount of outgassed water is significantly smaller throughout the evolution. As a result, after 4.5 Gyr, $P_\mathrm{H_2O}$ is only 43\% (for $X^\mathrm{H_2O}_{m,0}=1000$ ppm) and 32\% (for $X^\mathrm{H_2O}_{m,0}=500$ ppm) of the partial pressure of H$_2$O that would be achieved if all water extracted at the surface were outgassed into the atmosphere.

\begin{figure}[ht!]
	\centering
	\resizebox{0.85\hsize}{!}{\includegraphics{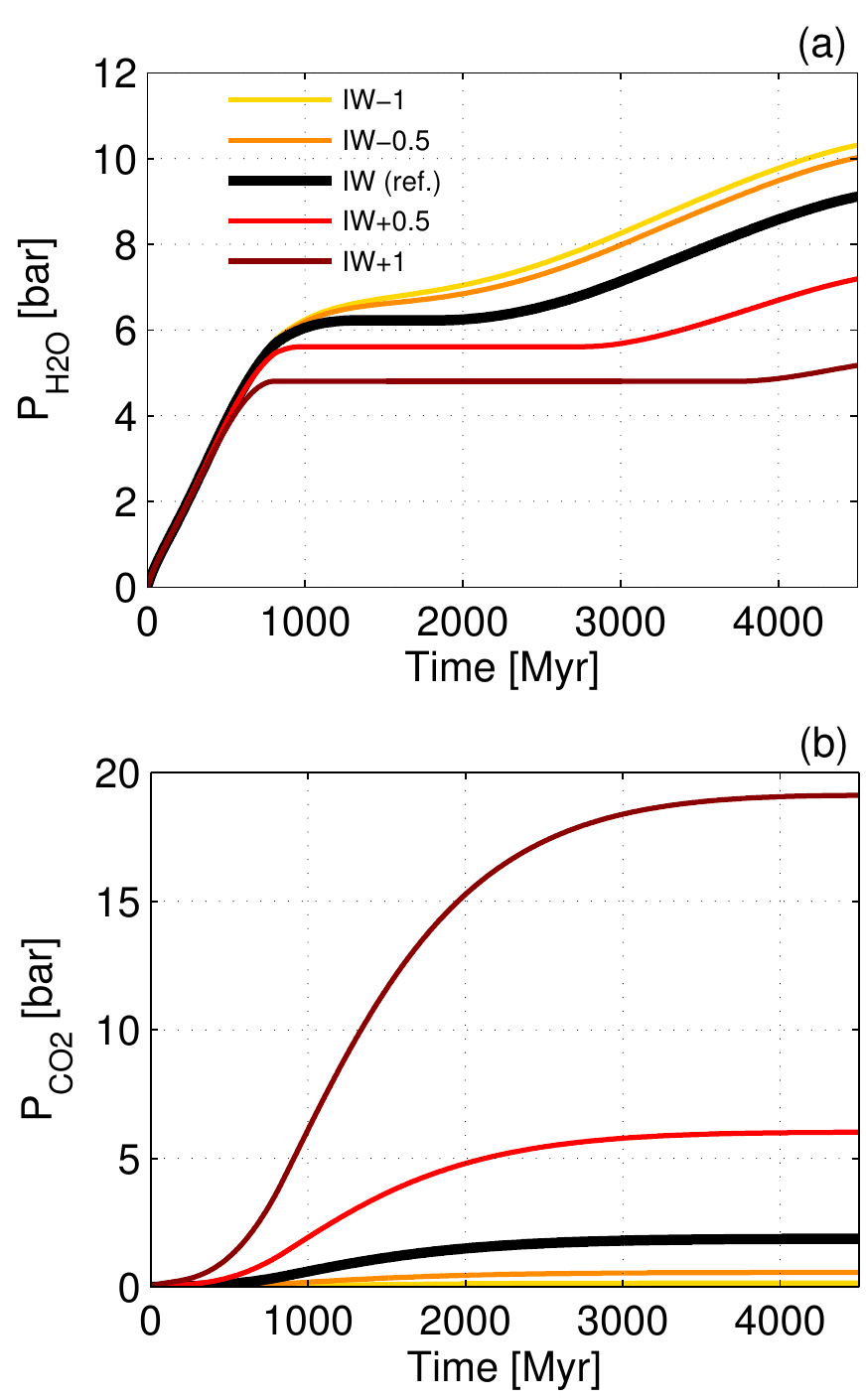}}
    \caption{Evolution of the partial pressure of outgassed H$_2$O (a) and CO$_2$ (b) for a model with $T_{m,0}=1700$ K, $X^\mathrm{H_2O}_{m,0}=500$ ppm and oxygen fugacities ranging from one log$_{10}$-unit below the IW buffer (IW-1) to one log$_{10}$-unit above it (IW+1). The black line indicates the reference model.} \label{fig_H2O_CO2_pressure}
\end{figure}

Water outgassing is suppressed even more strongly as more and more oxidizing conditions 
are assumed for the mantle. A higher oxygen fugacity leads in fact to a higher amount of outgassed CO$_2$ whose increasing pressure tends to prevent the concentration of H$_2$O in surface melts from exceeding its saturation level. Figure \ref{fig_H2O_CO2_pressure}a shows this effect for a model with $T_{m,0}=1700$ K, $X^\mathrm{H_2O}_{m,0}=500$ ppm, and $f_\mathrm{O_2}$ ranging from IW-1 to IW+1. The corresponding evolutions of the outgassed CO$_2$ are shown in Fig. \ref{fig_H2O_CO2_pressure}b. For example, for $f_\mathrm{O_2}$ at IW+1 (brown lines in Fig. \ref{fig_H2O_CO2_pressure}), about 3 bar CO$_2$ are outgassed within the first 750 Myr. This pressure is sufficient to completely stop the release of water whose partial pressure reaches $\sim\! 4.7$ bar at this time. Melts again become  supersaturated (see also Fig. \ref{fig_outgasevol}a) and additional 0.4 bar H$_2$O can be outgassed only after $\sim\! 4$ Gyr .

\begin{figure}[ht!]
	\centering
	\resizebox{0.85\hsize}{!}{\includegraphics{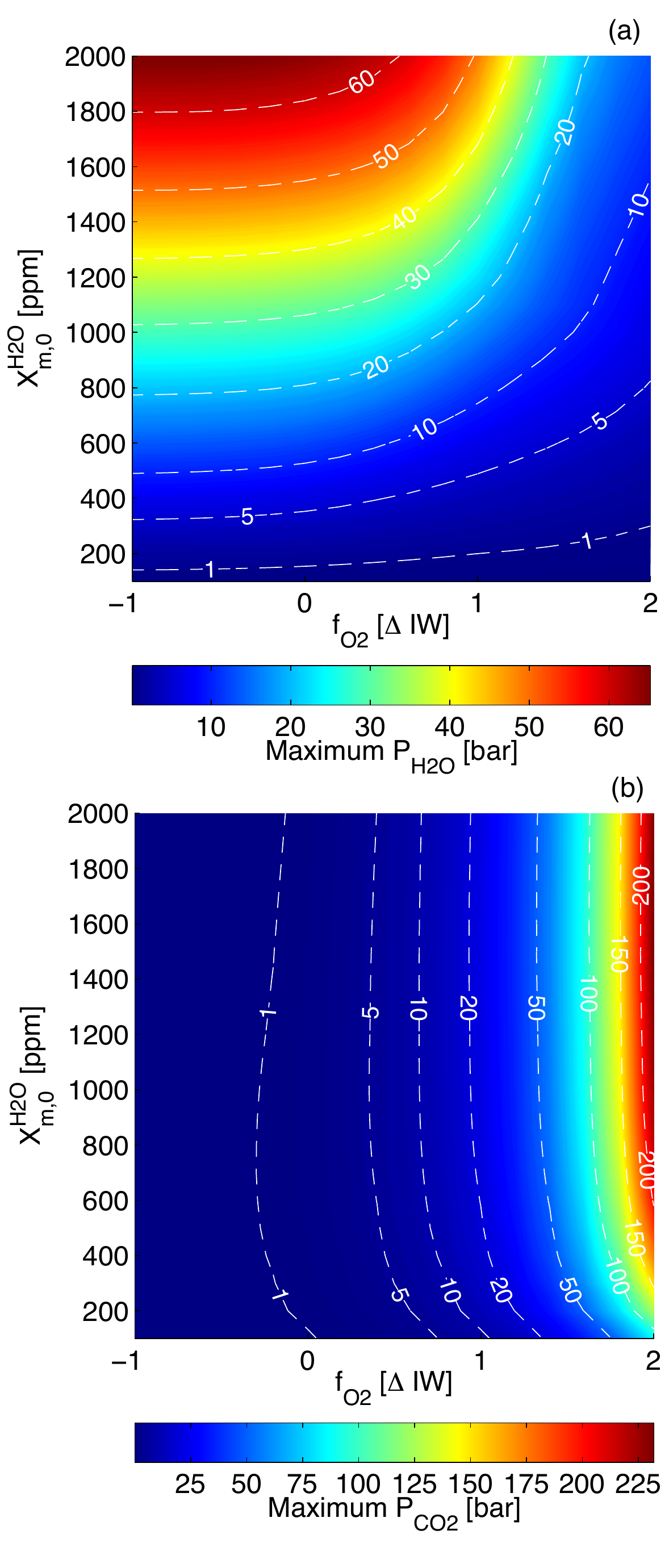}}
    \caption{Maximum partial pressures of outgassed H$_2$O (a) and CO$_2$ (b) after 4.5 Gyr of evolution as a function of the initial water concentration in the mantle ($X^\mathrm{H_2O}_{m,0}$) and oxygen fugacity ($f_{\text{O}_2}$) for an initial mantle temperature of 1700 K. } \label{fig_outgas_sum}
\end{figure}

In order to better appreciate the effects of the model parameters on the outgassing of the two volatiles, we
plotted in Fig. \ref{fig_outgas_sum} the partial pressure of H$_2$O  (\ref{fig_outgas_sum}a) and CO$_2$
(\ref{fig_outgas_sum}b) reached after 4.5 Gyr as a function of the initial water concentration between 100 and 2000
ppm and the oxygen fugacity between one log$_{10}$-unit below and two log$_{10}$-units above the IW buffer. As long
as the mantle oxygen fugacity is relatively low, the amount of outgassed CO$_2$ is very limited; for an oxygen
fugacity not higher than the IW buffer, the maximum partial pressure of CO$_2$ at the end of the evolution does not
exceed 1 bar (Fig. \ref{fig_outgas_sum}b). Under these conditions, the maximum partial pressure of water increases
linearly according to the initial water concentration in the mantle with $\sim\!1$ bar outgassed for
$X^\mathrm{H_2O}_{m,0}=100$ ppm and 65 bar for an extreme initial concentration of 2000 ppm (Fig.
\ref{fig_outgas_sum}a). The partial pressure of outgassed CO$_2$, however, increases roughly linearly with the
mantle oxygen fugacity, with tens to hundreds of bars of CO$_2$ at IW+1  and IW+2, respectively. Already for $f_\mathrm{O_2}$ between IW and IW+1, corresponding to the presence of a few bars of CO$_2$ in the atmosphere, $P_\mathrm{H_2O}$ increases with $X^\mathrm{H_2O}_{m,0}$ more slowly because of the increased solubility of water in surface lavas. For $f_{\text{O}_2}$ values higher than one log-unit above the IW buffer, the maximum H$_2$O pressure is limited to 10--15 bar at most, even for the highest values of  $X^\mathrm{H_2O}_{m,0}$. On the other hand, as shown in Fig. \ref{fig_outgas_sum}b, the final partial pressure of CO$_2$, which is much less soluble than water in basalts (Sect. \ref{sec_co2-outgas}), is largely determined by the choice of the oxygen fugacity. The initial concentration of water only affects significantly the amount of outgassed CO$_2$ for $X^\mathrm{H_2O}_{m,0}$ values below $\sim\! 400$ ppm. In contrast to the enrichment of heat sources and water in partial melts, which is calculated in dependence of a partition coefficient (Eq. \ref{eq_fracmelt}), the concentration of CO$_2$ is directly proportional to the melt fraction (Eq. \ref{eq_Xliqave}) and depends on the assumed oxygen fugacity (Eqs.  \ref{eq_XCO2a} and \ref{eq_XCO2b}). At low water concentrations the solidus is relatively high and melt fractions are small (Fig. \ref{fig_Tsol_eta}a). As a consequence, the effect of the latter on the extraction of CO$_2$ is more significant than that of $f_{\text{O}_2}$. For $X^\mathrm{H_2O}_{m,0} \gtrsim 400$ ppm, instead, melt fractions become less relevant and CO$_2$ outgassing is completely controlled by the oxygen fugacity.

\subsection{Atmospheric evolution} \label{sec_atmosevol}

\begin{figure}[ht!]
	\centering
	\resizebox{0.85\hsize}{!}{\includegraphics{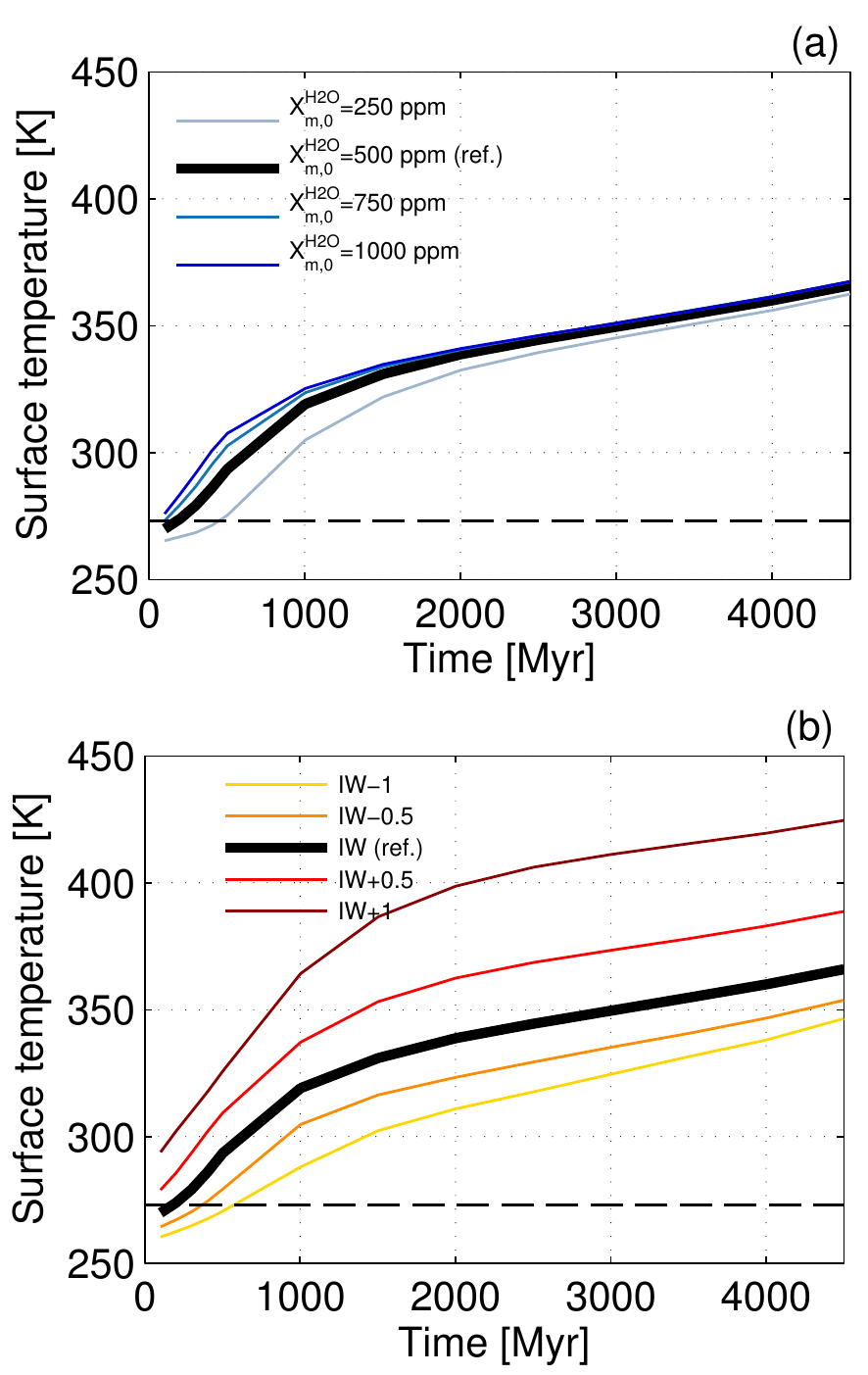}}
    \caption{Evolution of the surface temperature for (a) an oxygen fugacity at the IW buffer and different initial mantle concentrations of water (in different shades of blue), (b) an initial mantle concentration of water of 500\,ppm and different oxygen fugacities (in different shades of yellow and red) of an Earth-sized stagnant-lid planet located at 1\,au around the evolving Sun. The reference scenario ($T_{m,0}=1700$ K, $X_{m,0}^\mathrm{H_2O} = 500$, and $f_\mathrm{O_2}$ at the IW buffer) is indicated in black. The freezing point of water (273\,K) is indicated by the dashed black line in both panels. The temperatures in panel a correspond to the outgassing evolutions of H$_2$O and CO$_2$ plotted in Figs. \ref{fig_outgasevol}c and \ref{fig_outgasevol}d, while the temperatures of panel b correspond to the outgassing evolutions of Figs. \ref{fig_H2O_CO2_pressure}a and \ref{fig_H2O_CO2_pressure}b.} \label{fig_atmo_evolution}
\end{figure}

The interior evolution and resulting outgassing of CO$_2$ and H$_2$O into the planetary atmosphere lead to an evolution of the planetary climate. Since CO$_2$ and H$_2$O are important greenhouse gases, their abundance has a strong impact on the surface temperatures. Figure \ref{fig_atmo_evolution} shows the evolution of the surface temperatures of Earth-like stagnant-lid planets for different initial mantle concentrations of H$_2$O  (Fig.~\ref{fig_atmo_evolution}a) and oxygen fugacities (Fig.~\ref{fig_atmo_evolution}b). For these scenarios, the planet is located at an orbital distance of 1\,au around the evolving Sun. 

The stagnant-lid planets show habitable surface conditions over most of their history. Only during the early evolution up to 500\,Myr, some scenarios  -- with relatively low initial  mantle concentrations of water or low oxygen fugacities -- show temperatures below 273\,K.  We use the term habitable for conditions in which the global mean surface temperature is above the freezing point of water (273\,K) but still low enough to allow for liquid water on the surface of the planet. 

The surface temperature evolution is controlled by the outgassing of CO$_2$ and H$_2$O from the interior and the increase in solar luminosity. The H$_2$O outgassed from the interior serves as a water reservoir. The amount of water vapour in the atmosphere then depends on the surface temperature of the planet. As shown in Fig.~\ref{fig_atmo_evolution}a for an oxygen fugacity at the IW buffer, the initial mantle concentration of water has only a small effect on the atmospheric evolution since, at these temperatures, the water vapor concentrations in the atmosphere are lower than the outgassed water reservoir. For an initial mantle water concentration of 250\,ppm, however, the outgassing of CO$_2$ is much slower (see Fig.~\ref{fig_outgasevol}), which leads to a slower increase in surface temperatures. While at the lower temperatures of the atmosphere, which can be found during the early evolution, the CO$_2$ abundance is important, it shows a negligible effect at later  stages; the greenhouse effect of water vapor is so strong that the difference in CO$_2$ of about 0.7\,bar arising from the use of different initial values of the mantle water concentration (see Fig.~\ref{fig_outgasevol}d) does not exert a significant effect.

For different oxygen fugacities, the amount of CO$_2$ outgassed into the atmosphere varies much more significantly among the various scenarios (Fig.~\ref{fig_H2O_CO2_pressure}). As shown in Fig.~\ref{fig_atmo_evolution}b, this leads to a larger impact upon the evolution of the surface temperatures with temperature differences up to 75 K after 4500 Myr.
We performed atmospheric calculations for scenarios with oxygen fugacities up to IW+1, which leads to atmospheric CO$_2$ partial pressures of about 20 bar. For higher partial pressures, the model boundary condition of a zero downwelling infrared radiation at the top-of-the-atmosphere becomes invalid. For these higher partial pressures we experienced difficulties in reaching a converged solution in radiative equilibrium.

\begin{figure}[ht!]
	\centering
	\resizebox{\hsize}{!}{\includegraphics{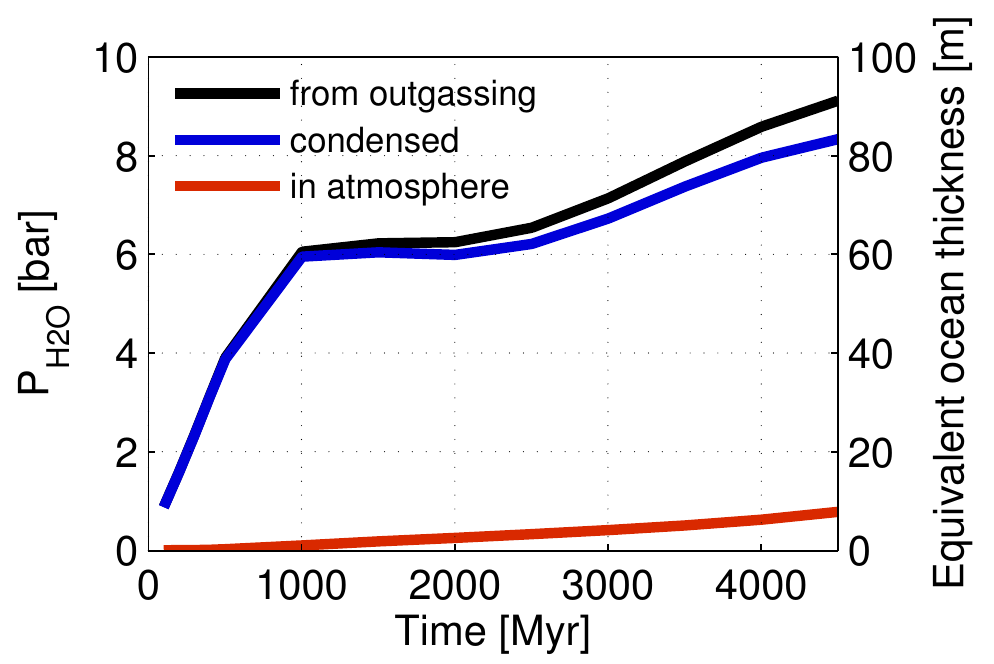}}
        \caption{Evolution of the partition of the outgassed H$_2$O (black) between atmosphere (red) and ocean (blue) for the reference scenario with an initial mantle concentration of water of 500\,ppm, an initial mantle temperature of 1700\,K, and an oxygen fugacity corresponding to the IW buffer.}
        \label{fig_water_evolution}
\end{figure}

Figure \ref{fig_water_evolution} shows the evolution of the partitioning of the water outgassed from the interior into the atmosphere and the ocean for the reference scenario with an initial mantle concentration of water of 500\,ppm, an initial mantle temperature of 1700\,K, and an oxygen fugacity at the IW buffer. Most of the water is stored in the  ocean and the atmospheric abundance defined via the saturation vapor pressure is comparatively low at the surface temperatures obtained for an orbital distance of 1\,au. For most of the evolution, the majority of the oceanic water can be liquid, except for the first 100\,Myr, in which surface temperatures below 273\,K are found and thus at least part of the oceanic water can be expected to be solid. After 4500\,Myr, a water amount of about 9 bar, which corresponds to about 3\% of an Earth ocean, would form an ocean on the stagnant-lid planet for this scenario, which corresponds to an equivalent depth of about 85\,m.

\begin{figure}[ht!]    
    \centering
    \resizebox{0.85\hsize}{!}{\includegraphics{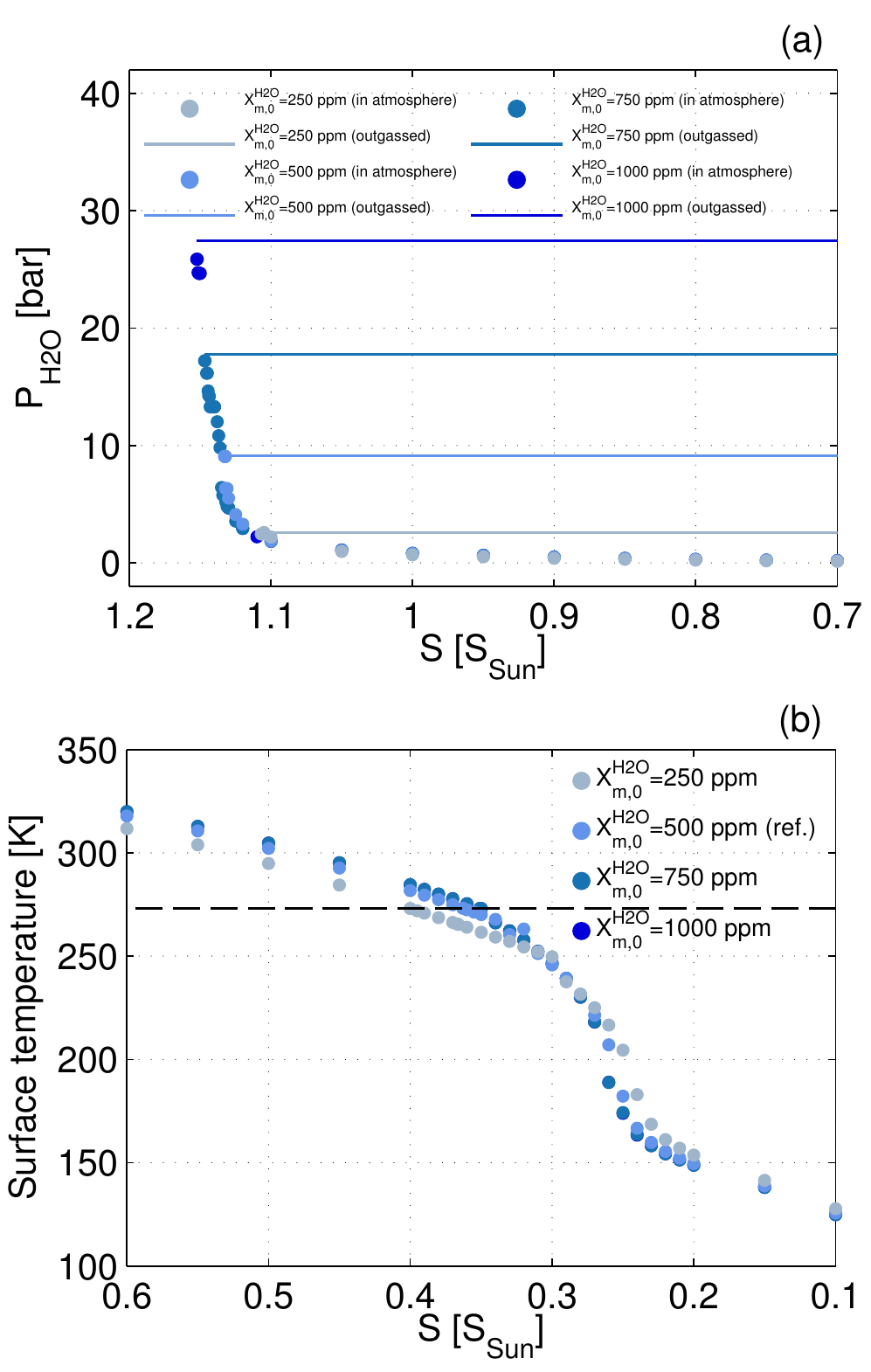}}
    \caption{Determination of the inner (a) and outer (b) HZ boundaries in solar irradiation $S$ via evaporation of the entire water reservoir ($P_{\mathrm{H_2O,out}}$), and the criteria of a minimum temperature of 273.15\,K (dashed black line in panel b) for an oxygen fugacity at the IW buffer and initial mantle water concentrations between 250 and 1000\,ppm (indicated in different shades of blue). Owing to similar behaviours of the atmospheres, the model results are overlaid.} \label{fig_determination_HZ}
\end{figure}

Figure \ref{fig_determination_HZ} shows how we determined the boundaries of the HZ for our stagnant-lid planets. For the inner boundary, the insolation has been varied to find the amount of irradiation at which the surface temperature is so high that the complete water reservoir outgassed from the interior is in its vapor phase according to the saturation vapor pressure of water (Fig.~\ref{fig_determination_HZ}a). It can be inferred that although the amount of water to be evaporated increases when the initial mantle concentration of water is increased, the difference in insolation to evaporate the water reservoir completely is small. At these temperatures, the water vapor feedback is very efficient, leading to a strong increase in surface temperature and atmospheric water content for a small increase in solar irradiation. 
For the determination of the inner boundary of the HZ, we use the  insolation where at least 90\% of the outgassed amount of H$_2$O is evaporated into the atmosphere. For the determination of the outer boundary of the HZ, the insolation was varied to find the irradiation that results in a global mean surface temperature of 273\,K for the given CO$_2$ amount outgassed from the interior. Fig.~\ref{fig_determination_HZ}b shows this procedure for different initial mantle water concentrations and an oxygen fugacity at the IW buffer. H$_2$O in the atmosphere is determined via the surface temperature, which is set to 273\,K, and therefore does not vary. 

\begin{figure}[ht!]
    \centering
    \resizebox{0.85\hsize}{!}{\includegraphics{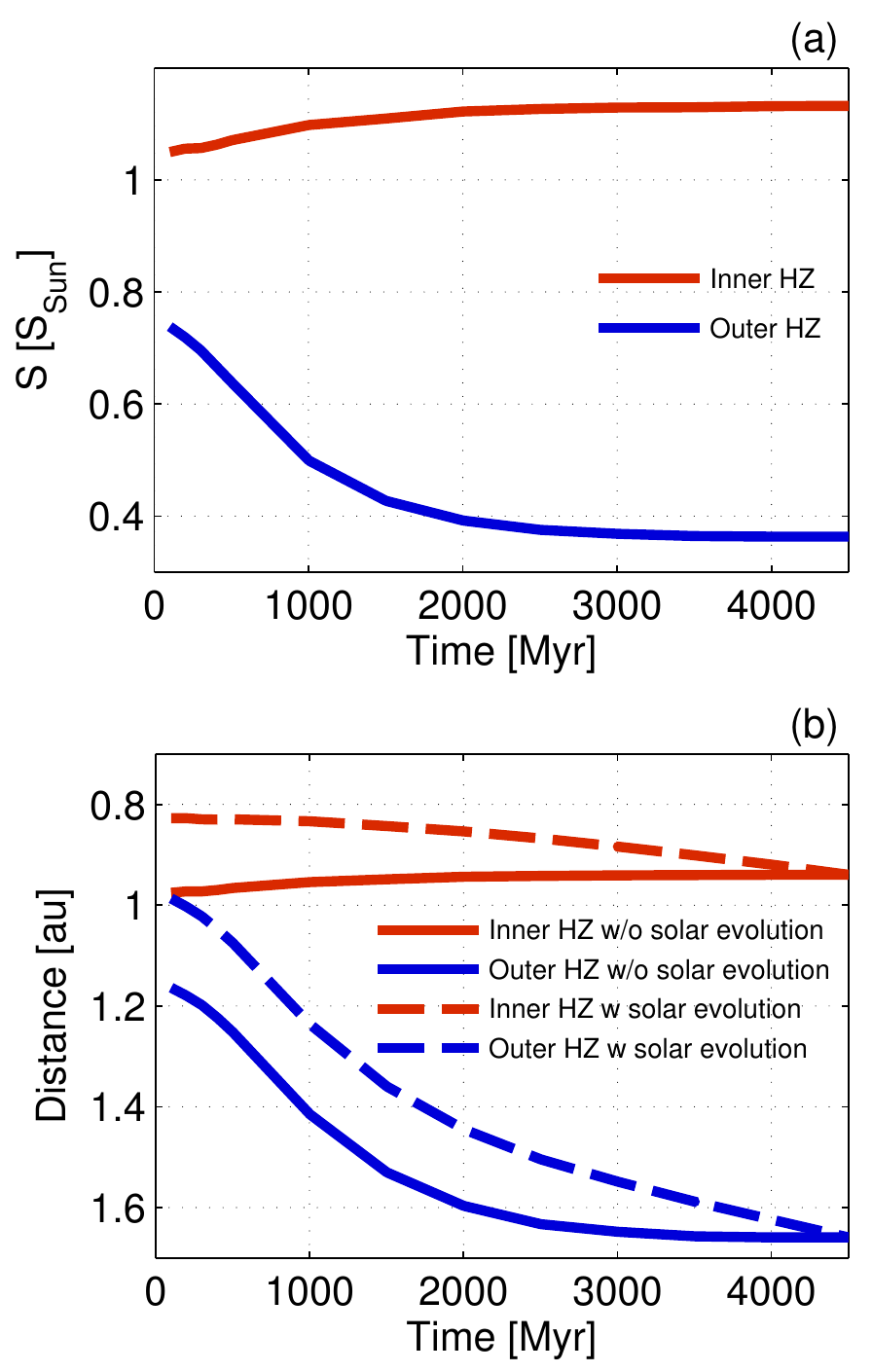}}
    \caption{Evolution of the inner (in red) and outer (in blue) boundary of the HZ in (a) unit of solar constants (S$_\text{Sun}$) and  
    (b) HZ boundary evolution in orbital distance (au) for the reference scenario. The impact of the increasing solar luminosity can be inferred from comparing the HZ distance evolution with (dashed) and without (solid) solar evolution. 
    } \label{fig_HZ_evolution}
\end{figure}

Figure \ref{fig_HZ_evolution} shows the evolution of the HZ boundaries for the reference scenario with an initial mantle water concentration of 500\,ppm and mantle temperature of 1700\,K. Figure \ref{fig_HZ_evolution}a shows the evolution of the inner (red) and outer (blue) boundary with time in units of the solar constant. This figure shows that the outer edge of the HZ strongly varies in time in response to the outgassing from the interior, mainly of CO$_2$, whose abundance increases during the evolution. The amount of water vapor in the atmosphere is constant since it is defined via the saturation vapor pressure at 273\,K. The inner edge of the HZ, expressed in total solar irradiation, shows a small increase with time during the early evolution, which corresponds to an increase in water outgassed from the interior (see Fig.~\ref{fig_p-effect}) since a larger water reservoir needs to be evaporated. At later evolutionary stages the inner boundary is found at a nearly constant insolation with time since the amount of water outgassed from the interior is  approximately constant.

Figure \ref{fig_HZ_evolution}b shows the orbital distances at which a planet would receive the insolation needed to reach the boundaries of the HZ. The solid lines show the boundaries when neglecting the solar evolution with time, while the dashed lines show the orbital distance including the fainter Sun during the early evolution \citep[following][]{gough1981}. When neglecting the solar evolution, the orbital distances of the HZ boundaries show a similar evolution as the HZ boundaries in terms of solar irradiation (Fig.~\ref{fig_HZ_evolution}). Including the solar luminosity evolution leads to an increase in orbital distance of the inner HZ with time as the Sun brightens. At the outer edge of the HZ, the increase in orbital distance with time due to the increase in CO$_2$ in the atmosphere is enhanced by the brightening Sun. However, the distances at the beginning of the evolution are shifted towards the Sun. The width of the HZ increases with time owing to the increase in atmospheric CO$_2$ and is larger when accounting for the solar evolution because of the $r^{-2}$ dependence of the stellar flux at the position of the planet.

\begin{figure}[ht!]
	\centering
	\resizebox{\hsize}{!}{\includegraphics{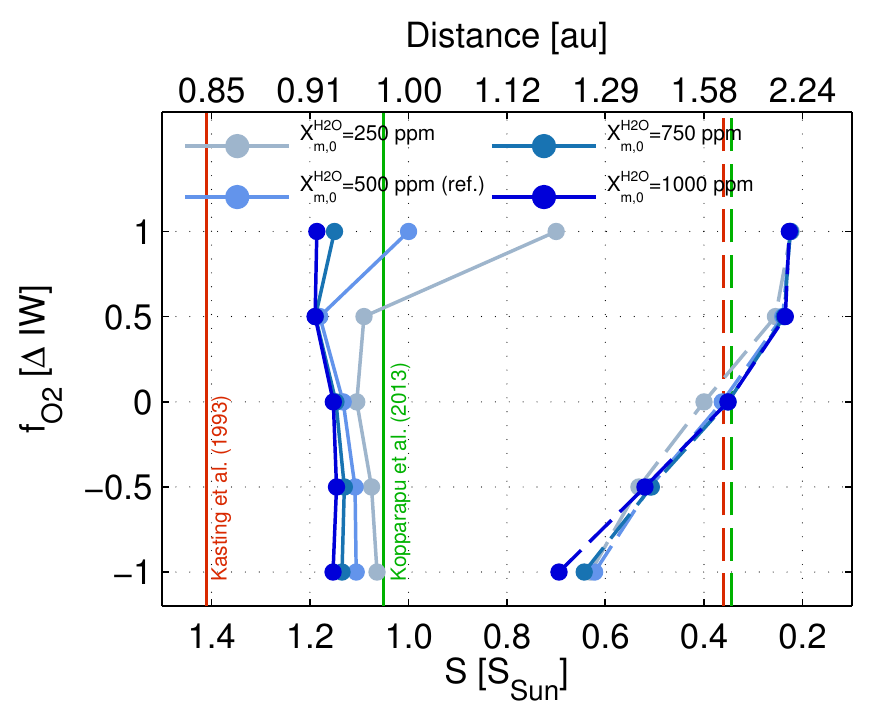}}
        \caption{Habitable zone boundaries for different oxygen fugacities (relative to the IW buffer) and initial water concentrations in the mantle (indicated in different shades of blue) in comparison to other HZ boundaries (\citet{kasting1993a} in red and \citet{kopparapu2013} in green). Solid and dashed lines refer to inner and outer HZ boundaries, respectively. 
        }\label{fig_HZ_over_IW}
\end{figure}

Figure \ref{fig_HZ_over_IW} shows the HZ boundaries after 4.5\,Gyr of evolution for different interior scenarios in comparison to the HZ boundaries as determined by \citet{kasting1993a} and \citet{kopparapu2013}.  
For low oxygen fugacities, the inner edge of the HZ is located at similar  insolations, with a slight shift towards the Sun until $f_{\text{O}_2}$ approaches a value of IW+0.5. At IW+1, and for low initial water concentrations, the inner boundary is located at smaller insolations. While the amount of CO$_2$ outgassed from the interior increases with the oxygen fugacity, the amount of water  decreases because of its increased solubility in surface lavas. This behaviour is therefore the result of the competing influences of H$_2$O and CO$_2$ as radiative gases, which are present in different abundances. 

The increase in insolation needed to reach the inner boundary of the HZ when moving from low to medium oxygen fugacities (from IW-1 to about IW or IW+0.5 depending on the initial water concentration in the mantle) is caused by an increase in planetary albedo as the amount of CO$_2$ increases, while the greenhouse effect in the atmosphere owing to CO$_2$ and H$_2$O is nearly constant (not shown). At an oxygen fugacity of IW+1, the planetary albedo still increases, but also the greenhouse effect of CO$_2$ becomes more efficient, as CO$_2$ rises from about 4.5 to 14.2\,bar for an initial mantle concentration of water of 250\,ppm upon increasing the oxygen fugacity from IW+0.5 to IW+1. This leads to a noticeable shift of the inner HZ away from the Sun because, already at less stellar irradiation, higher surface temperatures are reached as a result of the high amount of CO$_2$ in  the atmosphere. In addition, the amount of water released from the interior decreases from about 2.1 to 1.6\,bar for the same scenarios (for an initial water mantle concentration of 250\,ppm when moving form IW+0.5 to IW+1), hence less water is available that can be evaporated. 

At the outer boundary of the HZ, the amount of solar irradiation needed to keep the global mean surface temperature at the freezing point of water decreases with increasing oxygen fugacity, hence CO$_2$ outgassed from the interior, owing to the increasing greenhouse effect. At oxygen fugacities between IW+0.5 and IW+1, however, the decrease is much smaller, showing that about the same insolation is needed to obtain 273\,K for atmospheres containing about 10\,bar more CO$_2$. This is because not only the greenhouse effect is increased for increasing amounts of CO$_2$ but also scattering becomes more efficient. 

\section{Discussion}\label{sec_discussion}
%
\subsection{Stagnant-lid versus plate-tectonic planets: Interior and outgassing evolution} \label{sec_disc_ptsl}

There are a few studies on the long-term evolution of the habitability of the Earth and other planets in which calculations of CO$_2$ outgassing based on actual models of the thermal and melting history of the mantle have been used in combination with atmospheric models of varying complexity to predict the resulting climate \citep[e.g.][]{kadoya2014,kadoya2015,foley2016}. From the point of view of modelling of the interior, the most important difference between our approach and those above is that we assume our planet to operate in a stagnant-lid mode of convection rather than in the mobile-lid (or active-lid) mode realized through plate tectonics. On the one hand, in stagnant-lid bodies the presence of a thick, immobile lithosphere limits the production of partial melt, making this possible only at large depths where the solidus is relatively high. As the mantle cools and the lithosphere thickens, melting tends first to wane (Fig. \ref{fig_interiorevol}c) and then to vanish completely within a few billion years \citep{kite2009}. On the other hand, in a planet with an active lid as the Earth where thin tectonic plates form the lithosphere, melting can take place over a wide pressure range up to shallow depths of few kilometers beneath mid-ocean ridges with the consequence that outgassing can typically occur over the entire lifetime of the host star despite the cooling of the interior \citep{kite2009}.

In contrast to a plate-tectonic body, for a stagnant-lid body, volatiles cannot be cycled as efficiently as they are cycled, for example, in the water and carbon cycles of the Earth, even though our model allows for the possibility of some reintroduction of crustal -- and hence volatile-rich -- material into the mantle and therefore is not strictly a one-way street, at least not for water. Still, as Fig.~\ref{fig_interiorevol}d shows, there is a secular loss of water from the deep interior. The absence of subduction in our models has also an indirect influence on the chemistry of the interior that may feed back into the degassing behaviour and, in turn, the atmosphere. The recycling of water, which in the Earth is considered an oxidizing agent reintroduced into the mantle chiefly at subduction zones \citep[e.g.][]{McCammon05} is comparatively small in our models. Therefore, the evolution from a reducing environment towards a more oxidizing environment, which has been postulated at least for the upper mantle of the Earth \citep{frost2008}, would be strongly inhibited in a stagnant-lid planet. This justifies a posteriori our assumption that the oxygen fugacity can be held fixed at a rather low value. In the context of plate tectonics and subduction, this assumption would probably have to be dropped, although how and to which extent these processes would alter the oxygen fugacity of the mantle is poorly known. At any rate, an evolution towards more oxidizing conditions would imply that carbon would change from its reduced, bound form into the volatile form of CO$_2$ that can be degassed, as long as any carbon is available. In the absence of significant sinks of carbon at the surface, an enormous CO$_2$ pressure would build up in the atmosphere. This is also suggested by Fig.~\ref{fig_outgas_sum}b, although that diagram was not calculated for such a scenario. 

Our model of redox melting considerably simplifies the treatment of CO$_2$ extraction and outgassing, which could become extremely complex in a plate-tectonic setting. In thermal evolution models based on plate tectonics, the amount of outgassed CO$_2$ is typically treated independently of the mantle composition and simply related to the seafloor spreading rate and depth of melt generation \citep[e.g.][]{kadoya2015,foley2016}. Yet carbon in the upper mantle of the Earth is stored in a number of accessory phases rather than being bound to silicates as in the case of water \citep{dasgupta2010}. Together with the fact that under oxidizing conditions, the presence of carbon can dramatically reduce the solidus of domains of carbonated mantle \citep{dasgupta2006}, it is clear that modelling the melting and extraction of carbon under these conditions would be a challenging task, particularly when using 1D models that cannot account for lateral variations in the volatile distribution.

Thermal models based on plate-tectonic convection generally assume this mode to operate throughout the planetary evolution. Although the surface rock record of the Earth and numerical models suggest that some form of surface mobilization and recycling may have already been present during the Archean, subduction in the early Earth, if existent at all, was probably characterized by an episodic, intermittent behaviour \citep[see][and references therein]{vanhunen2012}. It is only during the past two to three billion years that modern-style subduction, consisting of a continuous, regular creation of new seafloor at mid-ocean ridges and of recycling of cold plates at subduction zones, is likely to have been active \citep[e.g.][]{condie2016}. The possibility that plate tectonics may have not operated uniformly over the evolution of the Earth, in particular that it could have followed an initial stagnant-lid state \citep[e.g.][]{korenaga2013,moore2013} and could even be a transient rather than an end-member state \citep{oneill2016} makes the modelling of long-term mantle melting and outgassing particularly difficult and subject to major uncertainties. Furthermore, the present understanding of the basic physical mechanisms that are ultimately responsible for the generation of plate tectonics is still incomplete \citep{bercovici2003a}. Since the solar system provides us with examples of bodies that do not show any evidence for present or past plate tectonics, our simplifying assumption of considering a stagnant-lid planet appears thus capable of delivering robust results.

Owing to the similarity of Earth and Venus in terms of size and composition and to the fact that the latter does not show evidence of plate tectonics today, it is also tempting to compare our outgassing results with the composition of the atmosphere of Venus; since we have limited our atmosphere calculations to temperatures below the critical point of water, we cannot reproduce Venus-like climates. At present, the atmosphere of Venus contains 92 bar of CO$_2$ and, with only 20 ppm of water vapour, is essentially dry; this is a figure that is actually not very surprising, at least as far as the amount of CO$_2$ is concerned. Earth and Venus are thought to have similar compositions \citep[e.g.][]{fegley2005}. But while on Venus CO$_2$ resides mostly in the atmosphere, on Earth it is largely contained in carbonate rocks. Indeed, as much as the equivalent of 60 bar of CO$_2$ is thought to be stored in carbonates on Earth \citep{kasting1988}, whose formation is strongly facilitated by the presence of water, which Venus lost during its early evolution \citep[see e.g.][]{kasting1984a,taylor2009}. As shown in Fig. \ref{fig_outgas_sum}b, by choosing relatively oxidizing conditions for the mantle ($f_\mathrm{O_2}$ between IW+1.5 and IW+2), it is not difficult to obtain a 90-bar present-day atmosphere, nearly independently of the initial water concentration of the mantle. The build-up of such a thick atmosphere would also prevent water from being released from surface lavas and would thus lead to a strongly water-depleted atmosphere. Our models generally predict CO$_2$ to be continuously outgassed during the evolution, albeit at a decreasing rate (Fig. \ref{fig_H2O_CO2_pressure}b). The abundance of CO$_2$ in the atmosphere of the Earth is buffered over geological timescales by surface weathering and the carbon--silicate cycle. Whether a comparable process able to buffer the CO$_2$ concentration over long timescales is also at play on Venus is unclear \citep{taylor2009}. Yet understanding whether such a process actually operates would help to clarify whether the surface pressure of Venus was acquired early and remained nearly stable over the evolution \citep{gillmann2014} or evolved substantially over time, which is the scenario predicted by our models. A direct comparison with Venus is also not straightforward because we assume that the stagnant-lid regime persists throughout the evolution. Venus, in contrast, with its young surface, is thought to have experienced a large-scale resurfacing event about 1 Gyr ago \citep{romeo2010}. This event, which replenished the ancient surface with new volcanic material, may also have been accompanied by a significant injection of volatiles into the atmosphere, whose extent, however, is difficult to constrain.

The question as to whether the atmospheric pressure of Venus has a primordial origin or evolved over time is also related to a fundamental assumption of our models, namely that the atmospheres that we considered are generated solely by secondary outgassing of H$_2$O and CO$_2$ from the interior and that any primordial atmosphere, whether accreted from the nebula or degassed by a magma ocean has been lost, except for the presence of 1 bar of N$_2$. The assumption of 1 bar N$_2$ can be motivated by a comparison of Venus and Earth, both of which hold about the same amount of N$_2$. Furthermore, this assumption allows for a better comparison with other habitability studies (Sect. \ref{sec_disc_habitability}) such as those performed by \citet{kasting1993a} and \citet{kopparapu2013}. At any rate, the impact of this choice is not very significant. We tested the effect of a lower amount of N$_2$ on the surface temperatures for our reference scenario ($X_{m,0}^\mathrm{H_2O}=500$ ppm and $f_\mathrm{O_2}$ at the IW buffer) after 4.5 Gyr and found that with a N$_2$ pressure of 0.1\,bar we obtain a surface temperature of 357.45\,K, while with only 0.01\,bar we obtain 356.40\,K, i.e. less than 10\,K lower than in the case with 1 bar N$_2$.

The neglect of a thick primordial atmosphere such as one in excess of tens to hundreds of bars of H$_2$O and CO$_2$
that could be generated by catastrophic outgassing of a magma ocean \citep[e.g.][]{elkinstanton2008,lebrun2013} has
a potentially much more significant influence on the outcomes of our models. In contrast to previous studies, no
matter whether based on plate-tectonic or stagnant-lid convection, our models account for the effect of growing
atmospheric pressure on the outgassing of H$_2$O and CO$_2$ (see Figs. \ref{fig_p-effect} and
\ref{fig_H2O_CO2_pressure}). Although with the kind of atmospheres that we have considered the outgassing of CO$_2$
is hardly influenced by this effect because of its low solubility (Fig. \ref{fig_solubilities}b), the outgassing of
H$_2$O, which on the contrary is much more soluble in basaltic lavas (Fig. \ref{fig_solubilities}a), can be strongly suppressed, particularly when the mantle oxygen fugacity is large enough for significant quantities of CO$_2$ to be released into the atmosphere. Indeed, as shown in Fig. \ref{fig_outgas_sum}a, depending on the actual amount of outgassed CO$_2$, this poses strict limits on the amount of water that can be brought to the surface and atmosphere via volcanism. Therefore, depending on the pressure of a primordial atmosphere and on its resilience to escape, volcanic outgassing of H$_2$O could be easily suppressed from the very beginning of the evolution, and subsequent outgassing of CO$_2$ could also be hindered.

\subsection{Habitability} \label{sec_disc_habitability}

The term ``habitability'' denotes in a broad sense conditions that are hospitable to life. Life as we know it has three basic requirements, namely (i) an energy source, (ii) a solvent, and (iii) the potential for molecular complexity \citep[see][]{horneck2006}. The classical HZ \citep[][]{huang1960,kasting1993a} refers to the annulus region around a star where a planet could support liquid water as a solvent on its surface. The wider, non-classical HZ however \citep[see e.g.][]{lammer2009} extends further out to include, for example icy moons beyond the snowline, which are widely accepted to have subsurface oceans.

In exoplanetary science, habitability is commonly defined using the classical HZ concept although clearly the existence of liquid water is not the only criterion. Numerous other potentially important factors have been discussed in the literature. These include (i) the availability of nutrients, for example in the form of the elements CHNOPS \citep[see e.g.][]{horneck2016}; (ii) the formation and maintenance of an atmosphere that enables the presence of liquid water \citep[][]{grenfell2010}; (iii) the existence of a magnetic field that could protect the surface from high-energy particles \citep[][]{lundin2007}; (iv) climate stabilization, possibly via the presence of a large moon \citep[][]{laskar1993}; (v) environmental diversity and climate-stabilizing feedbacks, for example due to the presence of plate tectonics which favours the carbonate--silicate cycle \citep[e.g.][]{korenaga2011, hoening2016}; (vi) planetary protection from impacts due to the presence of gas giants \citep[][]{horner2008}; and (vii) the properties of the central star \citep[][]{beech2011}. It should be noted that complex molecules such as proteins have a limited range of temperature over which they are stable. Most proteins can exist up to 50\textcelsius, while some can exist up to 120\textcelsius\ \citep[see e.g.][]{lineweaver2012}. A detailed review of the concept of habitability can be found in \citet{cockell2016}.

In this study we focused on the availability of liquid water as a criterion for habitability as used in the concept of the classical HZ. We used a mean surface temperature of 273\,K as the lowest habitable temperature. However, recent studies have shown that liquid water on a planetary surface is also possible for global mean temperatures below 273\,K with temperatures as low as $\sim\! 235$\,K since, locally, the surface temperatures are above 273 K and may still allow for liquid water \citep[e.g.][]{Charnay2013,Wolf2013,Kunze2014,Shields2014,Godolt2016}. 

At high temperatures, we have not defined a limit of habitability in terms of surface temperature, but rather in terms of water reservoir available from outgassing. We defined the upper limit for habitability at the stellar irradiation at which the entire water reservoir  outgassed from the interior would be in its vapor phase according to the assumption of phase equilibrium at the surface, as  previously done also by \citet{pollack1971}. Therefore this limit occurs at different surface temperatures. 
This approach differs from most of the definitions proposed in the literature. A hard limit used by \citet{kasting1993a} and \citet{kopparapu2013} is the runaway greenhouse limit. This limit defines the stellar irradiation at which a planet with a water reservoir of one Earth ocean and an Earth-like atmosphere does not allow for climate states with liquid water on the planetary surface as the water feedback tends to increase the surface temperatures and pressures above the critical point of water.  \citet{kasting1993a} also suggested another habitability limit, the water loss limit. It is defined at a stellar irradiation at which a water reservoir of one Earth ocean (270\,bar) may be lost to space within 4.5\,Gyr.  Water loss is thought to be efficient if stratospheric water vapor volume mixing ratios exceed a critical value of about 3$\times$10$^{-3}$ \citep[][]{kasting1993a}. This limit is estimated to occur at temperatures of about 340 to 350\,K \citep[e.g.][]{selsis2007, Wolf2015}. Some of our stagnant-lid scenarios show higher global mean surface temperatures at later stages of the evolution and may therefore lose their water reservoir, which is even smaller than one Earth ocean, via photodissociation of H$_2$O molecules and subsequent loss of hydrogen to space.
 
 The comparison of our model results with the HZ boundaries in Fig.~\ref{fig_HZ_over_IW} shows that the inner boundary of the HZ determined for the stagnant-lid scenarios lies almost for all cases in between the boundaries determined by \citet{kasting1993a} and \citet{kopparapu2013}. In Fig.~\ref{fig_HZ_over_IW} we show the inner boundary of the HZ that these two studies determined as the runaway greenhouse limit for an Earth-like planet with a water reservoir of one Earth ocean (270\,bar). The main difference between these two studies lies in the spectral databases and continua used to derive the radiative fluxes in the planetary atmosphere.  It has been shown that especially the treatment of the radiative properties of water vapor can lead to different results at the inner edge of the HZ \citep[][]{yang2016}.  Here we used yet another spectral database (HITEMP1995) and continuum assumptions different from those of \citet{kasting1993a} and \citet{kopparapu2013}.  Furthermore, we used a forward climate modelling approach, where we specify the stellar irradiation, the atmospheric pressure and composition (except for water vapor, which is calculated) and calculate the surface temperatures self-consistently, while \citet{kasting1993a} and \citet{kopparapu2013} apply inverse climate modelling.  In this approach, the temperature profile is fixed by specifying a surface temperature, following a dry or moist adiabat until a defined stratospheric temperature of, for example~200\,K is met.  For this temperature, the stellar irradiation needed to balance the outgoing infrared radiation is then derived. The 1D climate model used in this study would result in an insolation of about 1.32 S$_\mathrm{Sun}$ for the inner boundary of the HZ, when using the inverse climate modelling approach, specifying the surface temperature at about 647\,K (the critical point of water), a stratospheric temperature of 200\,K, a water reservoir of one Earth ocean, and deriving the insolation needed via the assumption of radiative energy balance at top of the atmosphere. Hence, the classical inner HZ boundary determined with our 1D model would lie in between the results of \cite{kasting1993a} (at 1.41 $S_\mathrm{Sun}$) and \cite{kopparapu2013} (at 1.06 $S_\mathrm{Sun}$). This result arises because we use a different database and continua. In comparison to this boundary considering an Earth ocean as water reservoir,  the inner boundary of the HZ of the stagnant-lid planets generally lies further away from the star the smaller the water reservoirs obtained via outgassing from the interior are.
 
 The assumption of a saturated atmosphere as made in our study and in \citet{kasting1993a} and \citet{kopparapu2013} has been questioned by recent 3D modelling studies \citep[see e.g.][]{leconte2013}. The impact of different relative humidities upon 1D climate modelling results and their comparison to 3D model results has been investigated by \citet{Godolt2016}. The assumption of a fully saturated atmosphere overestimates the surface temperatures at higher temperatures. Because of this the inner boundaries determined with this assumption are pessimistic. For lower relative humidities, the inner boundary of the HZ would move towards the star. For even smaller water reservoirs than those found here, the HZ could move inwards even further, as shown by, for example \citet{Abe2011}, \citet{Zsom2013}, or \citet{Leconte2013b}, since the greenhouse effect of water would be very low at the resulting low atmospheric volume mixing ratios.  
 
 We found that the inner edge of the HZ may indeed depend on the amount of CO$_2$ in the atmosphere. \citet{kasting1993a} found that the runaway greenhouse limit for a planet with a water reservoir of one Earth ocean does not depend on the amount of CO$_2$ in the atmosphere, while the water loss limit occurs at smaller orbital distances if the atmosphere contains more CO$_2$. A study by \citet{popp2016} also found that the inner boundary of the HZ as determined by water loss can be strongly influenced by the CO$_2$ content of the atmosphere. Since we obtain a much smaller water reservoir than one Earth ocean by outgassing from the interior of a stagnant-lid planet, the atmospheres at the inner edge of the HZ are not necessarily dominated by water vapor as is the case in the runaway greenhouse scenarios as modelled by \citet{kasting1993a}. Therefore, increasing the amount of CO$_2$ to a few tens of bars completely changes the mean composition of the atmospheres and can therefore also  have an impact on the planetary climate.
 
 The outer boundaries of the HZ determined for the stagnant-lid planets lie partly within and partly outside the HZ as determined by \citet{kasting1993a} and \citet{kopparapu2013} depending mainly on the oxygen fugacity. Our scenarios, which lie within the HZ, have lower partial pressures of CO$_2$ than those assumed by the other two studies. For higher oxygen fugacities, the outer boundary of the HZ lies outside the outer boundary determined by \citet{kasting1993a} and \citet{kopparapu2013}, which is surprising. We cannot infer a maximum greenhouse effect from our model calculations, which range up to CO$_2$ partial pressures of 22 bar.  At high partial pressures of CO$_2$, we found the outer edge of the HZ to be fairly constant. In contrast, when applying the inverse climate modelling approach used by \citet{kasting1993a} and \citet{kopparapu2013} and assuming a stratospheric temperature of 150K, we obtain a maximum greenhouse effect at CO$_2$ partial pressures of about 4\,bar. With this approach the outer boundary of the HZ would be located at 0.33S$_{\mathrm{Sun}}$, which is at a slightly lower insolation as for \cite{kasting1993a} (at 0.36S$_{\mathrm{Sun}}$) and \cite{kopparapu2013} (at 0.35S$_{\mathrm{Sun}}$) because of the effect of different databases and continua. Exploring the effect of inverse versus forward climate modelling upon the boundaries of the HZ in detail is beyond the scope of this paper. At the outer edge of the HZ, the assumption of a saturated atmosphere does not show a large impact as the water concentrations are still very low at 273\,K. It has been shown in other studies \citep[e.g.][]{pierrehumbert2011} that other greenhouse gases, such as molecular hydrogen, which may still be present from a primordial atmosphere, may expand the outer boundary of the HZ far beyond the boundary defined here.

In the present study we only considered a stagnant-lid planet around the Sun. The surface temperature and habitable  zone evolution of Earth-like stagnant-lid planets around other types of central stars can be expected to be different for two reasons. Firstly, the luminosity evolution is different for other stellar types, and secondly, the spectral stellar flux distribution is known to have a large impact on the planetary climate due to the wavelength-dependent absorption and scattering properties of the atmospheric compounds.

\subsection{Influence of surface processes}

One of the limitations of the models that we presented is the lack of consideration of surface processes that may alter the volatile budget predicted on the base of interior outgassing. 

Liquid water on the surface, whether in the form of a global or regional ocean, could provide a sink for the atmospheric CO$_2$. Such a sink is temperature dependent with a higher amount of CO$_2$ trapped in the ocean for colder temperatures (see e.g. \cite{pierrehumbert2010} and \cite{kitzmann2015}). Since we assume that all CO$_2$ outgassed from the interior resides within the atmosphere, at the inner boundary of the HZ we obtain the maximum heating, hence the smallest insolation, because at these temperatures some CO$_2$ could still be bound in the ocean. At the outer HZ boundary, part of the CO$_2$ would be dissolved in the ocean (more than at the inner edge due to the temperature dependence of the solubility), which would lead to lower CO$_2$ concentration in the atmosphere and hence probably a lower greenhouse effect. However, the ocean is smaller than the Earth ocean, hence less CO$_2$ would be dissolved. 

Apart from the ocean, there are other sinks for carbon and water that would have to be considered in a more detailed model. Among the most complex ones is probably the sequestration of carbon dioxides by carbonate minerals such as calcite (CaCO$_{3}$) \citep[e.g.][]{walker1981} and of water in various hydrated silicates. While in the modern Earth the biogenic production of CaCO$_{3}$ is most familiar and has been producing carbonate sediments for hundreds of millions of years, there are also abiogenic processes that may also operate in a planet that does not, or does not yet, host life. One such process of this type is the low-temperature ($<60$\textcelsius) alteration of the uppermost part of the ocean floor in the modern Earth, in which carbonates form from basaltic rock, but its quantitative significance is controversial \citep[e.g.][]{AlTe99}. It is nonetheless worthwhile considering because it does not require plate tectonics;~a carbon dioxide source and the existence of some sort of hydrothermal activity in basalt, which is a widespread rock type on other terrestrial planets without plate tectonics, should be sufficient. For instance, \citet{vBerk:etal12} modelled Mg--Fe-rich carbonates in the Comanche outcrop in Gusev crater on Mars, which was analysed by the Mars Exploration Rover Spirit, and found that one possible scenario for their formation involves a water-rich environment at temperatures around 280\,K and 0.5--2\,bar $P_{\mathrm{CO}_2}$, as may have prevailed during the Noachian. \citet{AlTe99} analysed borehole cores from various ocean drilling sites that probed young oceanic crust and observed a decrease of bound carbon with depth into the extrusive rock; most of the CO$_{2}$ is stored in the upper 600\,m. They found an average content of 0.214\,wt.\% CO$_{2}$ in the bulk rock, while \citet{Staudigel14} arrived at a total carbonate uptake of 0.355\,wt.\% and an uptake of 0.45\,wt.\% of crystal-bound water. A synopsis of data from the literature suggests that although young volcanic crustal rock may not take up a major part of this CO$_{2}$ and water, the oceanic crust remains an active sink as it ages and continues to react for some tens of millions of years \citep[e.g.][]{Staudigel14}. 

An important difference between the modern Earth and the stagnant-lid Earth-like planet considered in this study is that in the latter, there is no mechanism that transports carbonated or hydrated rock back into the mantle at a steady rate, thus completing the carbon and water cycles, although periodic crustal delamination may still operate.  Moreover, the mode of crust production is also different, as mid-ocean ridges would also be absent: the crust would be built by intrusions and, in a top--down manner rather than lateral addition, by extrusives. If the supply of fresh basalt ceases, for instance, because the crust becomes too thick to allow the ascent of lava to the surface or because the mantle becomes too cool, the crustal carbon and water sinks are not renewed and therefore are saturated after a certain period. In the atmospheric evolution, such a situation would result in a delay of the atmospheric accumulation of that volatile, as opposed to the suppressed or reduced accumulation expected in evolutions without a sink. In a planet with plate tectonics, the availability of carbon sinks would be even more important to prevent the build-up of a massive CO$_2$ atmosphere because the increase in oxygen fugacity may reinforce CO$_2$ degassing, as discussed above (see Sect. \ref{sec_disc_ptsl}).

The weathering of impact ejecta is closely related to the sequestration of carbon and water in normal crust. Early in the history of the terrestrial planets, the much higher impact flux has produced abundant fresh rock surfaces of mostly mafic to ultramafic chemistry that were available for weathering \citep{KosterVanGroos88,ZaSl02}. The latter authors have carried out model calculations for CO$_{2}$ that suggest that ejecta weathering could bind enough of it to suppress the greenhouse effect entirely during the first few hundred millions of years. The possible effects of impacts go beyond the production of reactive surfaces, however. Depending on the sizes, velocities, and compositions of impactors, the impact processes themselves may result either in the erosion or the replenishment of the atmosphere of the target planet \citep{Ahrens93,Pham:etal11,deNiem:etal12}. Moreover, very large impacts also affect the deeper interior and may trigger volcanic activity, which in turn affects the atmosphere \citep{Marc:etal16}. The effects especially of rare, large impacts depend strongly on the parameters of the specific event and cannot be predicted in a unique general manner.

\section{Conclusions}\label{sec_conclusions}
%
The upcoming generation of exoplanetary missions holds promise for detecting small rocky planets orbiting Sun-like stars in the HZ \citep{rauer2014}. The measurement of mass and radius alone does not well constrain whether these bodies have plate tectonics. We have thus studied the interior and climate evolution of an Earth-like planet operating in the stagnant-lid mode of convection, which presently characterizes all solid bodies of the solar system other than the Earth. 

We analysed the evolution of the surface temperature and boundaries of the HZ based solely on volcanic degassing of H$_2$O and CO$_2$ from the interior. The partial pressures of H$_2$O and CO$_2$ that accumulate over time in the atmosphere are controlled by the initial water concentration and by the oxygen fugacity of the mantle, respectively. As the total atmospheric pressure grows, the release of H$_2$O becomes more and more difficult because of its high solubility in surface lavas. After 4.5 Gyr, no more than a few tens of bars H$_2$O can be outgassed, even when assuming initial mantle concentrations in excess of 1000 to 2000 ppm. As a consequence, an Earth-sized ocean cannot be built up by secondary outgassing from the interior. On the other hand, CO$_2$, because it is much less soluble than water, can be outgassed throughout the evolution with partial pressures that depend on the assumed redox state of the mantle. While at reducing conditions ($f_\mathrm{O_2}$ from IW-1 to IW), only up to 1 bar CO$_2$ is outgassed, between 10 and 200 bar can be outgassed if the mantle is more oxidizing ($f_\mathrm{O_2}$ from IW+1 to IW+2).

At 1 au, for all cases that we analysed (i.e.  up to $f_\mathrm{O_2}$ at IW+1), the amount of H$_2$O and CO$_2$ outgassed from the interior leads to surface temperatures that allow for liquid water on the surface over nearly the entire evolution. 
A stagnant-lid Earth could then be habitable over geological timescales even though the obtained surface temperatures ($\sim\! 350$--420 K) may eventually lead to water loss to space and are hardly compatible with the limits for complex terrestrial life.  

The width of the HZ after 4.5 Gyr is controlled by the amount of outgassed CO$_2$, which in turn is determined by the mantle oxygen fugacity. For $f_\mathrm{O_2}$ at $\mathrm{IW}+1$ and assuming a relatively high mantle water concentration (e.g. 1000 ppm), for example, the HZ requires an insolation between $\sim\! 0.2$ and $1.2 S_\text{Sun}$. In contrast, for $f_\mathrm{O_2}$ at IW-1, the HZ is considerably thinner and an insolation between 0.7 and $1.15 S_\text{Sun}$ is necessary if the same H$_2$O concentration is assumed.

The outer edge of the HZ is mostly influenced by the outgassed CO$_2$, and is the farther away from the Sun the higher $f_\mathrm{O_2}$. The inner edge is characterized instead by a complex shape dependent on the amount of CO$_2$ and on that of H$_2$O. When the partial pressure of H$_2$O becomes too low in response to a CO$_2$ increase, in fact, the limited water reservoir can easily evaporate so that the inner edge of the HZ is significantly shifted away from the Sun. 

In conclusion, the joint modelling of interior evolution, volcanic outgassing, and accompanying climate is crucial for a robust determination of the habitability conditions on rocky exoplanets. 

\begin{acknowledgements}
We are grateful to Craig O'Neill for his constructive comments on an earlier version of this work.
N.Tosi acknowledges support from the Helmholtz Association (project VH-NG-1017) and from the German Research Foundation (DFG) through the Priority Programme 1833 ``Building a habitable Earth'' (grant TO 704/2-1). A.-C. Plesa acknowledges support from the Interuniversity Attraction Poles Programme initiated by the Belgian Science Policy Office through the Planet TOPERS alliance and from the DFG (SFB-TRR 170). T. Ruedas was supported by the DFG (grant Ru 1839/1-1 and SFB-TRR 170). This is TRR 170 Publication No. 29.
\end{acknowledgements}

\bibpunct{(}{)}{;}{a}{}{,}
\bibliographystyle{aa}
\bibliography{staglid-tr,references_nic}

\begin{thebibliography}{157}
\expandafter\ifx\csname natexlab\endcsname\relax\def\natexlab#1{#1}\fi

\bibitem[{{Abe} {et~al.}(2011){Abe}, {Abe-Ouchi}, {Sleep}, \&
  {Zahnle}}]{Abe2011}
{Abe}, Y., {Abe-Ouchi}, A., {Sleep}, N.~H., \& {Zahnle}, K.~J. 2011,
  Astrobiology, 11, 443

\bibitem[{Agee(2008)}]{agee2008}
Agee, C.~B. 2008, Phil. Trans. R. Soc. A, 366, 4239

\bibitem[{Ahrens(1993)}]{Ahrens93}
Ahrens, T.~J. 1993, Annu. Rev. Earth Planet. Sci., 21, 525

\bibitem[{Allen(1973)}]{allen1973}
Allen, C. 1973, Astrophysical Quantities (University of London: The Athlone
  Press)

\bibitem[{Alt \& Teagle(1999)}]{AlTe99}
Alt, J.~C. \& Teagle, D. A.~H. 1999, Geochim. Cosmochim. Acta, 63, 1527

\bibitem[{Armann \& Tackley(2012)}]{armann2012}
Armann, M. \& Tackley, P.~J. 2012, J. Geophys. Res.: Planets, 117

\bibitem[{Aubaud {et~al.}(2004)Aubaud, Hauri, \& Hirschmann}]{aubaud2004}
Aubaud, C., Hauri, E.~H., \& Hirschmann, M.~M. 2004, Geophys. Res. Lett., 31

\bibitem[{Aulbach \& Stagno(2016)}]{AuSt16}
Aulbach, S. \& Stagno, V. 2016, Geology, 44, 751

\bibitem[{Batalha(2014)}]{batalha2014}
Batalha, N.~M. 2014, Proc. Nat. Acad. Sci., 111, 12647

\bibitem[{{Beech}(2011)}]{beech2011}
{Beech}, M. 2011, J. R. Astron. Soc. Can., 105, 232

\bibitem[{Bercovici(2003)}]{bercovici2003a}
Bercovici, D. 2003, Earth Planet. Sci. Lett., 205, 107

\bibitem[{Bercovici \& Ricard(2003)}]{bercovici2003b}
Bercovici, D. \& Ricard, Y. 2003, Geophys. J. Int., 152, 581

\bibitem[{Blundy \& Wood(2003)}]{blundy2003}
Blundy, J. \& Wood, B. 2003, Earth Planet. Sci. Lett., 210, 383

\bibitem[{{Bucholtz}(1995)}]{bucholtz1995}
{Bucholtz}, A. 1995, Appl. Opt., 34, 2765

\bibitem[{Catling {et~al.}(2001)Catling, Zahnle, \& McKay}]{catling2001}
Catling, D.~C., Zahnle, K.~J., \& McKay, C.~P. 2001, Science, 293, 839

\bibitem[{{Charnay} {et~al.}(2013){Charnay}, {Forget}, {Wordsworth}, {Leconte},
  {Millour}, {Codron}, \& {Spiga}}]{Charnay2013}
{Charnay}, B., {Forget}, F., {Wordsworth}, R., {et~al.} 2013, J. Geophys. Res.:
  Atmospheres, 118, 10414

\bibitem[{Choblet \& Sotin(2000)}]{choblet2000}
Choblet, G. \& Sotin, C. 2000, Phys. Earth Planet. Inter., 119, 321

\bibitem[{Christensen(1984)}]{christensen1984}
Christensen, U.~R. 1984, Geophys. J. R. Astr. Soc., 77, 343

\bibitem[{Clough {et~al.}(1989)Clough, Kneizys, \& Davies}]{clough1989}
Clough, S., Kneizys, F., \& Davies, R. 1989, Atmos. Res., 23, 229

\bibitem[{{Cockell} {et~al.}(2016){Cockell}, {Bush}, {Bryce}, {Direito},
  {Fox-Powell}, {Harrison}, {Lammer}, {Landenmark}, {Martin-Torres},
  {Nicholson}, {Noack}, {O'Malley-James}, {Payler}, {Rushby}, {Samuels},
  {Schwendner}, {Wadsworth}, \& {Zorzano}}]{cockell2016}
{Cockell}, C.~S., {Bush}, T., {Bryce}, C., {et~al.} 2016, Astrobiology, 16, 89

\bibitem[{Condie(2016)}]{condie2016}
Condie, K.~C. 2016, Geoscience Frontiers

\bibitem[{Costa {et~al.}(2009)Costa, Caricchi, \& Bagsdassarov}]{costa2009}
Costa, A., Caricchi, L., \& Bagsdassarov, N. 2009, Geochem. Geophys. Geosyst.,
  10

\bibitem[{Dasgupta \& Hirschmann(2006)}]{dasgupta2006}
Dasgupta, R. \& Hirschmann, M.~M. 2006, Nature, 440, 659

\bibitem[{Dasgupta \& Hirschmann(2010)}]{dasgupta2010}
Dasgupta, R. \& Hirschmann, M.~M. 2010, Earth Planet. Sci. Lett., 298, 1

\bibitem[{Davaille \& Jaupart(1993)}]{davaille1993}
Davaille, A. \& Jaupart, C. 1993, J. Fluid Mech., 253, 141

\bibitem[{de~Niem {et~al.}(2012)de~Niem, K{\"u}hrt, Morbidelli, \&
  Motschmann}]{deNiem:etal12}
de~Niem, D., K{\"u}hrt, E., Morbidelli, A., \& Motschmann, U. 2012, Icarus,
  221, 495

\bibitem[{Deschamps \& Sotin(2001)}]{deschamps2001}
Deschamps, F. \& Sotin, C. 2001, J. Geophys. Res.: Planets, 106, 5107

\bibitem[{{Edl{\'e}n}(1966)}]{edlen1966}
{Edl{\'e}n}, B. 1966, Metrologia, 2, 71

\bibitem[{Elkins-Tanton(2008)}]{elkinstanton2008}
Elkins-Tanton, L. 2008, Earth Planet. Sci. Lett., 271, 181

\bibitem[{Fegley(2005)}]{fegley2005}
Fegley, B. 2005, in Treatise on Geochemistry, ed. A.~Davis (Elsevier), 127--148

\bibitem[{Foley \& Driscoll(2016)}]{foley2016}
Foley, B.~J. \& Driscoll, P.~E. 2016, Geochem. Geophys. Geosyst., 17, 1885–

\bibitem[{Frost \& McCammon(2008)}]{frost2008}
Frost, D.~J. \& McCammon, C.~A. 2008, Annu. Rev. Earth Planet. Sci., 36, 389

\bibitem[{Gaillard \& Scaillet(2014)}]{gaillard2014}
Gaillard, F. \& Scaillet, B. 2014, Earth Planet. Sci. Lett., 403, 307

\bibitem[{Gerya(2014)}]{gerya2014}
Gerya, T. 2014, Gondwana Research, 25, 442

\bibitem[{Gillmann \& Tackley(2014)}]{gillmann2014}
Gillmann, C. \& Tackley, P. 2014, J. Geophys. Res.: Planets, 119, 1189

\bibitem[{{Godolt} {et~al.}(2016){Godolt}, {Grenfell}, {Kitzmann}, {Kunze},
  {Langematz}, {Patzer}, {Rauer}, \& {Stracke}}]{Godolt2016}
{Godolt}, M., {Grenfell}, J.~L., {Kitzmann}, D., {et~al.} 2016, \aap, 592, A36

\bibitem[{{Goody} \& {Yung}(1989)}]{goody1989}
{Goody}, R.~M. \& {Yung}, Y.~L. 1989, {Atmospheric radiation : theoretical
  basis} (Oxford University Press, Inc)

\bibitem[{{Gough}(1981)}]{gough1981}
{Gough}, D.~O. 1981, Solar Physics, 74, 21

\bibitem[{Grasset \& Parmentier(1998)}]{grasset1998}
Grasset, O. \& Parmentier, E. 1998, J. Geophys. Res., 103, 18171

\bibitem[{{Grenfell} {et~al.}(2010){Grenfell}, {Rauer}, {Selsis},
  {Kaltenegger}, {Beichman}, {Danchi}, {Eiroa}, {Fridlund}, {Henning},
  {Herbst}, {Lammer}, {L{\'e}ger}, {Liseau}, {Lunine}, {Paresce}, {Penny},
  {Quirrenbach}, {R{\"o}ttgering}, {Schneider}, {Stam}, {Tinetti}, \&
  {White}}]{grenfell2010}
{Grenfell}, J.~L., {Rauer}, H., {Selsis}, F., {et~al.} 2010, Astrobiology, 10,
  77

\bibitem[{Grott \& Breuer(2010)}]{grott2010}
Grott, M. \& Breuer, D. 2010, J. Geophys. Res., 115

\bibitem[{Grott {et~al.}(2011{\natexlab{a}})Grott, Breuer, M., \&
  Laneuville}]{grott2011b}
Grott, M., Breuer, D., M., \& Laneuville. 2011{\natexlab{a}}, Earth Planet.
  Sci. Lett., 307, 135

\bibitem[{Grott {et~al.}(2011{\natexlab{b}})Grott, Morschhauser, Breuer, \&
  Hauber}]{grott2011}
Grott, M., Morschhauser, A., Breuer, D., \& Hauber, E. 2011{\natexlab{b}},
  Earth Planet. Sci. Lett., 308, 391

\bibitem[{{Gueymard}(2004)}]{gueymard2004}
{Gueymard}, C. 2004, Solar Energy, 76, 423

\bibitem[{Herd {et~al.}(2002)Herd, Borg, Jones, \& Papike}]{herd2002}
Herd, C.~D., Borg, L.~E., Jones, J.~H., \& Papike, J.~J. 2002, Geochim.
  Cosmochim. Acta, 66, 2025

\bibitem[{Hirschmann {et~al.}(2005)Hirschmann, Aubaud, \&
  Withers}]{hirschmann2005}
Hirschmann, M.~M., Aubaud, C., \& Withers, A.~C. 2005, Earth Planet. Sci.
  Lett., 236, 167

\bibitem[{Hirschmann \& Withers(2008)}]{hirschmann2008}
Hirschmann, M.~M. \& Withers, A.~C. 2008, Earth Planet. Sci. Lett., 270, 147

\bibitem[{Hirth \& Kohlstedt(2003)}]{hirth2003}
Hirth, G. \& Kohlstedt, D. 2003, in Inside the subduction Factory, ed.
  J.~Eiler, Geophysical Monograph Series (Wiley), 83--105

\bibitem[{Holloway(1998)}]{holloway1998}
Holloway, J.~R. 1998, Chem. Geol., 147, 89

\bibitem[{{H{\"o}ning} \& {Spohn}(2016)}]{hoening2016}
{H{\"o}ning}, D. \& {Spohn}, T. 2016, Phys. Earth Planet. Inter., 255, 27

\bibitem[{{Horneck}(2006)}]{horneck2006}
{Horneck}, G. 2006, in Reviews in Modern Astronomy, ed. S.~{Roeser}, Vol.~19,
  215

\bibitem[{{Horneck} {et~al.}(2016){Horneck}, {Walter}, {Westall}, {Grenfell},
  {Martin}, {Gomez}, {Leuko}, {Lee}, {Onofri}, {Tsiganis}, {Saladino},
  {Pilat-Lohinger}, {Palomba}, {Harrison}, {Rull}, {Muller}, {Strazzulla},
  {Brucato}, {Rettberg}, \& {Capria}}]{horneck2016}
{Horneck}, G., {Walter}, N., {Westall}, F., {et~al.} 2016, Astrobiology, 16,
  201

\bibitem[{{Horner} \& {Jones}(2008)}]{horner2008}
{Horner}, J. \& {Jones}, B.~W. 2008, Int. J. Astrobiol., 7, 251

\bibitem[{{Huang}(1960)}]{huang1960}
{Huang}, S.~S. 1960, Scientific American, 202, 55

\bibitem[{{Ingersoll}(1969)}]{ingersoll1969}
{Ingersoll}, A.~P. 1969, J. Atmos. Sci., 26, 1191

\bibitem[{Jambon \& Zimmermann(1990)}]{jambon1990}
Jambon, A. \& Zimmermann, J.~L. 1990, Earth Planet. Sci. Lett., 101, 323

\bibitem[{Kadoya \& Tajika(2014)}]{kadoya2014}
Kadoya, S. \& Tajika, E. 2014, \apj, 790, 107

\bibitem[{Kadoya \& Tajika(2015)}]{kadoya2015}
Kadoya, S. \& Tajika, E. 2015, \apjl, 815, L7

\bibitem[{{Kasting}(1988)}]{kasting1988}
{Kasting}, J.~F. 1988, Icarus, 74, 472

\bibitem[{{Kasting}(1991)}]{kasting1991}
{Kasting}, J.~F. 1991, Icarus, 94, 1

\bibitem[{Kasting {et~al.}(1993{\natexlab{a}})Kasting, Eggler, \&
  Raeburn}]{kasting1993b}
Kasting, J.~F., Eggler, D.~H., \& Raeburn, S.~P. 1993{\natexlab{a}}, J. Geol.,
  245

\bibitem[{{Kasting} {et~al.}(1984{\natexlab{a}}){Kasting}, {Pollack}, \&
  {Ackerman}}]{kasting1984a}
{Kasting}, J.~F., {Pollack}, J.~B., \& {Ackerman}, T.~P. 1984{\natexlab{a}},
  Icarus, 57, 335

\bibitem[{{Kasting} {et~al.}(1984{\natexlab{b}}){Kasting}, {Pollack}, \&
  {Crisp}}]{kasting1984b}
{Kasting}, J.~F., {Pollack}, J.~B., \& {Crisp}, D. 1984{\natexlab{b}}, J.
  Atmos. Chem., 1, 403

\bibitem[{Kasting {et~al.}(1993{\natexlab{b}})Kasting, Whitmire, \&
  Reynolds}]{kasting1993a}
Kasting, J.~F., Whitmire, D.~P., \& Reynolds, R.~T. 1993{\natexlab{b}}, Icarus,
  101, 108

\bibitem[{Katz {et~al.}(2003)Katz, Spiegelman, \& Langmuir}]{katz2003}
Katz, R., Spiegelman, M., \& Langmuir, C. 2003, Geochem. Geophys. Geosyst., 4

\bibitem[{Kite {et~al.}(2009)Kite, Manga, \& Gaidos}]{kite2009}
Kite, E.~S., Manga, M., \& Gaidos, E. 2009, Astrophys. J., 700, 1732

\bibitem[{{Kitzmann} {et~al.}(2015){Kitzmann}, {Alibert}, {Godolt}, {Grenfell},
  {Heng}, {Patzer}, {Rauer}, {Stracke}, \& {von Paris}}]{kitzmann2015}
{Kitzmann}, D., {Alibert}, Y., {Godolt}, M., {et~al.} 2015, \mnras, 452, 3752

\bibitem[{{Kopparapu} {et~al.}(2013){Kopparapu}, {Ramirez}, {Kasting}, {Eymet},
  {Robinson}, {Mahadevan}, {Terrien}, {Domagal-Goldman}, {Meadows}, \&
  {Deshpande}}]{kopparapu2013}
{Kopparapu}, R.~K., {Ramirez}, R., {Kasting}, J.~F., {et~al.} 2013, \apj, 765

\bibitem[{Kopparapu {et~al.}(2014)Kopparapu, Ramirez, SchottelKotte, Kasting,
  Domagal-Goldman, \& Eymet}]{kopparapu2014}
Kopparapu, R.~K., Ramirez, R.~M., SchottelKotte, J., {et~al.} 2014, Astrophys.
  J. Lett., 787, L29

\bibitem[{{Korenaga}(2011)}]{korenaga2011}
{Korenaga}, J. 2011, J. Geophys. Res.: Solid Earth, 116, B12403

\bibitem[{Korenaga(2013)}]{korenaga2013}
Korenaga, J. 2013, Annu. Rev. Earth Planet. Sci., 41, 117

\bibitem[{Koster Van~Groos(1988)}]{KosterVanGroos88}
Koster Van~Groos, A.~F. 1988, J. Geophys. Res., 93, 8952

\bibitem[{Kunze {et~al.}(2014)Kunze, Godolt, Langematz, Grenfell,
  Hamann-Reinus, \& Rauer}]{Kunze2014}
Kunze, M., Godolt, M., Langematz, U., {et~al.} 2014, Planet. Space Sci., 98, 77

\bibitem[{{Lammer} {et~al.}(2009){Lammer}, {Bredeh{\"o}ft}, {Coustenis},
  {Khodachenko}, {Kaltenegger}, {Grasset}, {Prieur}, {Raulin}, {Ehrenfreund},
  {Yamauchi}, {Wahlund}, {Grie{\ss}meier}, {Stangl}, {Cockell}, {Kulikov},
  {Grenfell}, \& {Rauer}}]{lammer2009}
{Lammer}, H., {Bredeh{\"o}ft}, J.~H., {Coustenis}, A., {et~al.} 2009, The
  Astronomy and Astrophysics Review, 17, 181

\bibitem[{{Laskar} {et~al.}(1993){Laskar}, {Joutel}, \& {Robutel}}]{laskar1993}
{Laskar}, J., {Joutel}, F., \& {Robutel}, P. 1993, Nature, 361, 615

\bibitem[{Lebrun {et~al.}(2013)Lebrun, Massol, Chassefi{\`e}re, Davaille,
  Marcq, Sarda, Leblanc, \& Brandeis}]{lebrun2013}
Lebrun, T., Massol, H., Chassefi{\`e}re, E., {et~al.} 2013, J. Geophys. Res.,
  118

\bibitem[{{Leconte} {et~al.}(2013{\natexlab{a}}){Leconte}, {Forget}, {Charnay},
  {Wordsworth}, \& {Pottier}}]{leconte2013}
{Leconte}, J., {Forget}, F., {Charnay}, B., {Wordsworth}, R., \& {Pottier}, A.
  2013{\natexlab{a}}, Nature, 504, 268

\bibitem[{{Leconte} {et~al.}(2013{\natexlab{b}}){Leconte}, {Forget}, {Charnay},
  {Wordsworth}, {Selsis}, {Millour}, \& {Spiga}}]{Leconte2013b}
{Leconte}, J., {Forget}, F., {Charnay}, B., {et~al.} 2013{\natexlab{b}}, \aap,
  554, A69

\bibitem[{L{\'e}cuyer \& Ricard(1999)}]{lecuyer1999}
L{\'e}cuyer, C. \& Ricard, Y. 1999, Earth Planet. Sci. Lett., 165, 197

\bibitem[{Li {et~al.}(2007)Li, Wentzcovitch, Weidner, \& Silva}]{li2007}
Li, L., Wentzcovitch, R.~M., Weidner, D.~J., \& Silva, C. S.~D. 2007, J.
  Geophys. Res., 112

\bibitem[{{Lineweaver} \& {Chopra}(2012)}]{lineweaver2012}
{Lineweaver}, C.~H. \& {Chopra}, A. 2012, Annu. Rev. Earth Planet. Sci., 40,
  597

\bibitem[{{Lundin} {et~al.}(2007){Lundin}, {Lammer}, \& {Ribas}}]{lundin2007}
{Lundin}, R., {Lammer}, H., \& {Ribas}, I. 2007, Space Sci. Rev., 129, 245

\bibitem[{Maal\o{e}(2004)}]{maaloe2004}
Maal\o{e}, S. 2004, Mineral. Petrol., 81, 1

\bibitem[{{Manabe} \& {Wetherald}(1967)}]{manabe1967}
{Manabe}, S. \& {Wetherald}, R.~T. 1967, J. Atmos. Sci., 24, 241

\bibitem[{Marchi {et~al.}(2016)Marchi, Black, Elkins-Tanton, \&
  Bottke}]{Marc:etal16}
Marchi, S., Black, B.~A., Elkins-Tanton, L.~T., \& Bottke, W.~F. 2016, Earth
  Planet. Sci. Lett., 449, 96

\bibitem[{{Marshall} \& {Smith}(1990)}]{marshall1990}
{Marshall}, B.~R. \& {Smith}, R.~C. 1990, Applied Optics, 29, 71

\bibitem[{McCammon(2005)}]{McCammon05}
McCammon, C.~A. 2005, in Earth's Deep Mantle: Structure, Composition, and
  Evolution, ed. R.~D. van~der Hilst, J.~D. Bass, J.~Matas, \& J.~Trampert,
  Geophysical Monograph No. 160 (American Geophysical Union), 219--240

\bibitem[{McDonough \& Sun(1995)}]{mcdonough1995}
McDonough, W.~F. \& Sun, S.-S. 1995, Chem. Geol., 120, 223

\bibitem[{McSween {et~al.}(2003)McSween, Grove, \& Wyatt}]{mcsween2003}
McSween, H.~Y., Grove, T.~L., \& Wyatt, M.~B. 2003, Journal of Geophysical
  Research: Planets, 108

\bibitem[{McSween {et~al.}(2009)McSween, Taylor, \& Wyatt}]{mcsween2009}
McSween, H.~Y., Taylor, G.~J., \& Wyatt, M.~B. 2009, Science, 324, 736

\bibitem[{{Mlawer} {et~al.}(1997){Mlawer}, {Taubman}, {Brown}, {Iacono}, \&
  {Clough}}]{mlawer1997}
{Mlawer}, E.~J., {Taubman}, S.~J., {Brown}, P.~D., {Iacono}, M.~J., \&
  {Clough}, S.~A. 1997, J. Geophys. Res., 102, 16663

\bibitem[{Moore \& Webb(2013)}]{moore2013}
Moore, W.~B. \& Webb, A.~G. 2013, Nature, 501, 501

\bibitem[{Morschhauser {et~al.}(2011)Morschhauser, Grott, \&
  Breuer}]{morschhauser2011}
Morschhauser, A., Grott, M., \& Breuer, D. 2011, Icarus, 212, 541

\bibitem[{Newman \& Lowenstern(2002)}]{newman2002}
Newman, S. \& Lowenstern, J.~B. 2002, Comput. Geosci., 28, 597

\bibitem[{Noack \& Breuer(2014)}]{noack2014b}
Noack, L. \& Breuer, D. 2014, Planet. Space Sci., 98, 41

\bibitem[{Noack {et~al.}(2014)Noack, Godolt, von Paris, Plesa, Stracke, Breuer,
  \& Rauer}]{noack2014a}
Noack, L., Godolt, M., von Paris, P., {et~al.} 2014, Planet. Space Sci., 98, 14

\bibitem[{O'Neill \& Lenardic(2007)}]{oneill2007c}
O'Neill, C. \& Lenardic, A. 2007, Geophys. Res. Lett., 34

\bibitem[{O’Neill {et~al.}(2016)O’Neill, Lenardic, Weller, Moresi,
  Quenette, \& Zhang}]{oneill2016}
O’Neill, C., Lenardic, A., Weller, M., {et~al.} 2016, Phys. Earth Planet.
  Inter., 255, 80

\bibitem[{O’Rourke \& Korenaga(2012)}]{orourke2012}
O’Rourke, J.~G. \& Korenaga, J. 2012, Icarus, 221, 1043

\bibitem[{{Pavlov} {et~al.}(2000){Pavlov}, {Kasting}, {Brown}, {Rages}, \&
  {Freedman}}]{pavlov2000}
{Pavlov}, A.~A., {Kasting}, J.~F., {Brown}, L.~L., {Rages}, K.~A., \&
  {Freedman}, R. 2000, J. Geophys. Res., 105, 11981

\bibitem[{Pham {et~al.}(2011)Pham, Karatekin, \& Dehant}]{Pham:etal11}
Pham, L. B.~S., Karatekin, {\"O}., \& Dehant, V. 2011, Planet. Space Sci., 59,
  1087

\bibitem[{{Pierrehumbert}(2010)}]{pierrehumbert2010}
{Pierrehumbert}, R. 2010, {Principles of Planetary Climate} (Cambridge Univ.
  Press)

\bibitem[{{Pierrehumbert} \& {Gaidos}(2011)}]{pierrehumbert2011}
{Pierrehumbert}, R. \& {Gaidos}, E. 2011, \apjl, 734, L13

\bibitem[{Plesa {et~al.}(2015)Plesa, Tosi, Grott, \& Breuer}]{plesa2015}
Plesa, A.-C., Tosi, N., Grott, M., \& Breuer, D. 2015, J. Geophys. Res.:
  Planets, 120, 995

\bibitem[{{Pollack}(1971)}]{pollack1971}
{Pollack}, J.~B. 1971, Icarus, 14, 295

\bibitem[{{Popp} {et~al.}(2016){Popp}, {Schmidt}, \& {Marotzke}}]{popp2016}
{Popp}, M., {Schmidt}, H., \& {Marotzke}, J. 2016, Nature Commun., 7, 10627

\bibitem[{Rauer {et~al.}(2014)}]{rauer2014}
Rauer, H. {et~al.} 2014, Experimental Astronomy, 38, 249

\bibitem[{Raymond {et~al.}(2004)Raymond, Quinn, \& Lunine}]{raymond2004}
Raymond, S.~N., Quinn, T., \& Lunine, J.~I. 2004, Icarus, 168, 1

\bibitem[{Reese {et~al.}(2005)Reese, Solomatov, \& Baumgardner}]{reese2005}
Reese, C., Solomatov, V., \& Baumgardner, J. 2005, Phys. Earth Planet. Inter.,
  149, 361

\bibitem[{Righter {et~al.}(2006)Righter, Drake, \& Scott}]{righter2006}
Righter, K., Drake, M., \& Scott, E. 2006, in Meteorites and the early solar
  system II, ed. D.~S. Lauretta \& H.~Y. McSween (Tucson: University of Arizona
  Press), 803--828

\bibitem[{Romeo \& Turcotte(2010)}]{romeo2010}
Romeo, I. \& Turcotte, D.~L. 2010, Planet. Space Sci., 58, 1374

\bibitem[{{Rothman} {et~al.}(1992){Rothman}, {Gamache}, {Tipping}, {Rinsland},
  {Smith}, {Benner}, {Devi}, {Flaud}, {Camy-Peyret}, \& {Perrin}}]{rothman1992}
{Rothman}, L.~S., {Gamache}, R.~R., {Tipping}, R.~H., {et~al.} 1992, J. Quant.
  Spectrosc. Radiat. Transf., 48, 469

\bibitem[{{Rothman} {et~al.}(2009){Rothman}, {Gordon}, {Barbe}, {Benner},
  {Bernath}, {Birk}, {Boudon}, {Brown}, {Campargue}, {Champion}, {Chance},
  {Coudert}, {Dana}, {Devi}, {Fally}, {Flaud}, {Gamache}, {Goldman},
  {Jacquemart}, {Kleiner}, {Lacome}, {Lafferty}, {Mandin}, {Massie},
  {Mikhailenko}, {Miller}, {Moazzen-Ahmadi}, {Naumenko}, {Nikitin}, {Orphal},
  {Perevalov}, {Perrin}, {Predoi-Cross}, {Rinsland}, {Rotger}, {{\v S}ime{\v
  c}kov{\'a}}, {Smith}, {Sung}, {Tashkun}, {Tennyson}, {Toth}, {Vandaele}, \&
  {Vander Auwera}}]{rothman2009}
{Rothman}, L.~S., {Gordon}, I.~E., {Barbe}, A., {et~al.} 2009, Journal of
  Quantitative Spectroscopy and Radiative Transfer, 110, 533

\bibitem[{{Rothman} {et~al.}(1995){Rothman}, {Wattson}, {Gamache}, {Schroeder},
  \& {McCann}}]{rothman1995}
{Rothman}, L.~S., {Wattson}, R.~B., {Gamache}, R., {Schroeder}, J.~W., \&
  {McCann}, A. 1995, in Society of Photo-Optical Instrumentation Engineers
  (SPIE) Conference Series, Vol. 2471, Society of Photo-Optical Instrumentation
  Engineers (SPIE) Conference Series, ed. {J.~C.~Dainty}, 105--111

\bibitem[{Saal {et~al.}(2002)Saal, Hauri, Langmuir, \& Perfit}]{saal2002}
Saal, A.~E., Hauri, E.~H., Langmuir, C.~H., \& Perfit, M.~R. 2002, Nature, 419,
  451

\bibitem[{{Schreier} \& {B{\"o}ttger}(2003)}]{schreier2003}
{Schreier}, F. \& {B{\"o}ttger}, U. 2003, Atmospheric and Oceanic Optics, 16,
  262

\bibitem[{{Schreier} \& {Schimpf}(2001)}]{schreier2001}
{Schreier}, F. \& {Schimpf}, B. 2001, in IRS 2000: Current Problems in
  Atmospheric Radiation (A. Deepak)

\bibitem[{{Segura} {et~al.}(2003){Segura}, {Krelove}, {Kasting}, {Sommerlatt},
  {Meadows}, {Crisp}, {Cohen}, \& {Mlawer}}]{segura2003}
{Segura}, A., {Krelove}, K., {Kasting}, J.~F., {et~al.} 2003, Astrobiology, 3,
  689

\bibitem[{{Selsis} {et~al.}(2007){Selsis}, {Kasting}, {Levrard}, {Paillet},
  {Ribas}, \& {Delfosse}}]{selsis2007}
{Selsis}, F., {Kasting}, J.~F., {Levrard}, B., {et~al.} 2007, \aap, 476, 1373

\bibitem[{{Shields} {et~al.}(2014){Shields}, {Bitz}, {Meadows}, {Joshi}, \&
  {Robinson}}]{Shields2014}
{Shields}, A.~L., {Bitz}, C.~M., {Meadows}, V.~S., {Joshi}, M.~M., \&
  {Robinson}, T.~D. 2014, \apjl, 785, L9

\bibitem[{Smrekar \& Sotin(2012)}]{smrekar2012}
Smrekar, S.~E. \& Sotin, C. 2012, Icarus, 217, 510

\bibitem[{Solomatov(1995)}]{solomatov1995}
Solomatov, V. 1995, Phys. Fluids, 7, 266

\bibitem[{Southam {et~al.}(2015)Southam, Westall, \& Spohn}]{southam2015}
Southam, G., Westall, F., \& Spohn, T. 2015, in Treatise on geophysics, vol.
  10, ed. T.~Spohn (Amsterdam: Elsevier), 473--486

\bibitem[{Spohn(1990)}]{spohn1991}
Spohn, T. 1990, Icarus, 90, 1222

\bibitem[{Stagno {et~al.}(2013)Stagno, Ojwang, McCammon, \& Frost}]{stagno2013}
Stagno, V., Ojwang, D.~O., McCammon, C.~A., \& Frost, D.~J. 2013, Nature, 493,
  84

\bibitem[{Stamenkovi{\'c} {et~al.}(2012)Stamenkovi{\'c}, Noack, Breuer, \&
  Spohn}]{stamenkovic2012}
Stamenkovi{\'c}, V., Noack, L., Breuer, D., \& Spohn, T. 2012, \apj, 748, 41

\bibitem[{Staudigel(2014)}]{Staudigel14}
Staudigel, H. 2014, in Treatise on Geochemistry, Vol.~4, The Crust, 2nd edn.,
  ed. R.~L. Rudnick (Elsevier), 583--606

\bibitem[{Stein {et~al.}(2013)Stein, Lowman, \& Hansen}]{stein2013}
Stein, C., Lowman, J., \& Hansen, U. 2013, Earth Planet. Sci. Lett., 361, 448

\bibitem[{Tackley(2000)}]{tackley2000a}
Tackley, P. 2000, Geochem. Geophys. Geosyst., 1

\bibitem[{Tajika(2007)}]{tajika2007}
Tajika, E. 2007, Earth Planets Space, 59, 293

\bibitem[{Taylor \& Grinspoon(2009)}]{taylor2009}
Taylor, F. \& Grinspoon, D. 2009, J. Geophys. Res.: Planets, 114

\bibitem[{{Toon} {et~al.}(1989){Toon}, {McKay}, {Ackerman}, \&
  {Santhanam}}]{toon1989}
{Toon}, O.~B., {McKay}, C.~P., {Ackerman}, T.~P., \& {Santhanam}, K. 1989, J.
  Geophys. Res., 941, 16287

\bibitem[{Tosi {et~al.}(2013)Tosi, Grott, Plesa, \& Breuer}]{tosi2013c}
Tosi, N., Grott, M., Plesa, A.-C., \& Breuer, D. 2013, J. Geophys. Res.
  Planets, 118, 2474

\bibitem[{Tozer(1967)}]{tozer1967}
Tozer, D. 1967, in The Earth's mantle, ed. T.~Gaskell (New York: Academic
  Press), 327--353

\bibitem[{Tuff {et~al.}(2013)Tuff, Wade, \& Wood}]{tuff2013}
Tuff, J., Wade, J., \& Wood, B.~J. 2013, Nature, 498, 342

\bibitem[{Turcotte(1993)}]{turcotte1993}
Turcotte, D.~L. 1993, J. Geophys. Res: Planets, 98, 17061

\bibitem[{Turcotte \& Schubert(2002)}]{turcotte2002}
Turcotte, D.~L. \& Schubert, G. 2002, Geodynamics (Cambridge: Cambridge
  University Press), 456 p.

\bibitem[{Valencia {et~al.}(2007)Valencia, O’connell, \&
  Sasselov}]{valencia2007}
Valencia, D., O’connell, R.~J., \& Sasselov, D.~D. 2007, \apjl, 670, L45

\bibitem[{van Berk {et~al.}(2012)van Berk, Fu, \& Ilger}]{vBerk:etal12}
van Berk, W., Fu, Y., \& Ilger, J.-M. 2012, J. Geophys. Res., 117

\bibitem[{van Heck \& Tackley(2011)}]{vanheck2011}
van Heck, H. \& Tackley, P. 2011, Earth Planet. Sci. Lett., 310, 252

\bibitem[{{van~Hunen} \& Moyen(2012)}]{vanhunen2012}
{van~Hunen}, J. \& Moyen, J.-F. 2012, Annu. Rev. Earth Planet. Sci., 40, 195

\bibitem[{{van Hunen} \& {van den Berg}(2008)}]{vanhunen2008}
{van Hunen}, J. \& {van den Berg}, A. 2008, Lithos, 103, 217

\bibitem[{{Vardavas} \& {Carver}(1984)}]{vardavas1984}
{Vardavas}, I.~M. \& {Carver}, J.~H. 1984, Planet. Space Sci., 32, 1307

\bibitem[{{von Paris}(2010)}]{paris2010a}
{von Paris}, P. 2010, PhD thesis, Technische Universit\"at Berlin

\bibitem[{{von Paris} {et~al.}(2010){von Paris}, {Gebauer}, {Godolt},
  {Grenfell}, {Hedelt}, {Kitzmann}, {Patzer}, {Rauer}, \&
  {Stracke}}]{paris2010b}
{von Paris}, P., {Gebauer}, S., {Godolt}, M., {et~al.} 2010, \aap, 522, A23+

\bibitem[{{von Paris} {et~al.}(2013){von Paris}, {Grenfell}, {Rauer}, \&
  {Stock}}]{vonParis2013b}
{von Paris}, P., {Grenfell}, J.~L., {Rauer}, H., \& {Stock}, J.~W. 2013,
  Planet. Space Sci., 82, 149

\bibitem[{{von Paris} {et~al.}(2008){von Paris}, {Rauer}, {Lee Grenfell},
  {Patzer}, {Hedelt}, {Stracke}, {Trautmann}, \& {Schreier}}]{paris2008}
{von Paris}, P., {Rauer}, H., {Lee Grenfell}, J., {et~al.} 2008, Planet. Space
  Sci., 56, 1244

\bibitem[{Wadhwa(2008)}]{wadhwa2008}
Wadhwa, M. 2008, Rev. Mineral. and Geochem., 68, 493

\bibitem[{Walker {et~al.}(1981)Walker, Hays, \& Kasting}]{walker1981}
Walker, J. C.~G., Hays, P.~B., \& Kasting, J.~F. 1981, J. Geophys. Res.:
  Oceans, 86, 9776

\bibitem[{White {et~al.}(2006)White, Crisp, \& Spera}]{white2006}
White, S.~M., Crisp, J.~A., \& Spera, F.~J. 2006, Geochem. Geophys. Geosyst., 7

\bibitem[{{Wiscombe} \& {Evans}(1977)}]{wiscombe1977}
{Wiscombe}, W.~J. \& {Evans}, J.~W. 1977, J. Comput. Phys., 24, 416

\bibitem[{{Wolf} \& {Toon}(2013)}]{Wolf2013}
{Wolf}, E.~T. \& {Toon}, O.~B. 2013, Astrobiology, 13, 656

\bibitem[{Wolf \& Toon(2015)}]{Wolf2015}
Wolf, E.~T. \& Toon, O.~B. 2015, J. Geophys. Res.: Atmospheres, 120, 5775,
  2015JD023302

\bibitem[{{Wordsworth} \& {Pierrehumbert}(2013)}]{wordsworth2013}
{Wordsworth}, R.~D. \& {Pierrehumbert}, R.~T. 2013, \apj, 778, 154

\bibitem[{{Yang} {et~al.}(2016){Yang}, {Leconte}, {Wolf}, {Goldblatt}, {Feldl},
  {Merlis}, {Wang}, {Koll}, {Ding}, {Forget}, \& {Abbot}}]{yang2016}
{Yang}, J., {Leconte}, J., {Wolf}, E.~T., {et~al.} 2016, \apj, 826, 222

\bibitem[{Zahnle \& Sleep(2002)}]{ZaSl02}
Zahnle, K. \& Sleep, N.~H. 2002, in The Early Earth: Physical, Chemical and
  Biological Development, Special Publication No. 199 (Geological Society of
  London), 231--257

\bibitem[{{Zsom} {et~al.}(2013){Zsom}, {Seager}, {de Wit}, \&
  {Stamenkovi{\'c}}}]{Zsom2013}
{Zsom}, A., {Seager}, S., {de Wit}, J., \& {Stamenkovi{\'c}}, V. 2013, \apj,
  778, 109

\end{thebibliography}

\end{document}